\documentclass[11pt,english,a4paper]{article}
\usepackage[top=1.5in,bottom=1.5in,left=1in,right=1in,heightrounded,marginparwidth=1.5in,marginparsep=1em]{geometry}
  
\usepackage[T1]{fontenc}
\usepackage{babel}

\usepackage[hyphens]{url}
\usepackage[hidelinks]{hyperref}
\hypersetup{breaklinks=true}

\usepackage{natbib}
\usepackage{amsmath}
\usepackage{amsfonts}
\usepackage{amssymb}

\usepackage{graphicx}
\usepackage{graphics}
\usepackage{placeins}
\usepackage{float}
\usepackage{chngcntr}
\usepackage{subcaption}

\usepackage{calc}
\usepackage{enumitem}

\usepackage{adjustbox}
\usepackage{tikz}
\usetikzlibrary{shapes.geometric,arrows,arrows.meta}
\usetikzlibrary{automata,fit,positioning}
\tikzset{%
  >={Latex[width=2mm,length=2mm]},
            base/.style = {rectangle, rounded corners, draw=black,
                           minimum width=4cm, minimum height=1cm,
                           text centered, font=\sffamily},
  start/.style = {base, fill=white!30},
       block/.style = {base, fill=white!30},
}

\usepackage[ruled,vlined]{algorithm2e}

\usepackage{marginnote}

\newcommand{\rzero}{ \operatorname{\text{R}_S} }

\usepackage{authblk}
\author[1]{Kevin Kunzmann\protect\footnote{\texttt{kevin.kunzmann@mrc-bsu.cam.ac.uk}}} 
\author[1]{Camilla Lingj{\ae}rde}
\author[1]{Sheila Bird}
\author[1]{Sylvia Richardson}
\affil[1]{MRC Biostatistics Unit, University of Cambridge}

\title{The `how' matters: A simulation-based assessment of the potential contributions of LFD tests for school reopening in England}
\date{\today}

\begin{document}

\maketitle

\begin{abstract}
During Covid-19 outbreaks, school closures are employed as part of governments' non-pharmaceutical interventions around the world to reduce the number of contacts and keep the reproduction number below 1.
Yet, prolonged school closures have profound negative
impact on the future opportunities of pupils, particularly from disadvantaged backgrounds,
as well as additional economic and social impacts by preventing their parents from returning to work.
Data on Covid-19 in children are sparse and policy frameworks are evolving quickly. 
We compare a set of potential policies to accompany the reopening of schools by means of an agent-based simulation tool.
The policies and scenarios we model reflect the public discussion and government guidelines in early March 2021 in England before the planned nationwide reopening of schools on the 8th of March. 
A point of particular interest is the potential contribution of a more wide-spread use of
screening tests based on lateral flow devices. 
We compare policies both with respect to their potential to contain new outbreaks of Covid-19 in schools and the proportion of schooldays lost due to isolation of pupils.
We find that regular asymptomatic screening of the whole school as an addition to a policy built around
isolation of symptomatic pupils and their closest contacts is beneficial across a wide range of scenarios, including when screening tests with relatively low test sensitivity are used.
Multiple screening tests per week bring only small additional benefits in some scenarios. 
These findings remain valid when test compliance is not enforced although the effectiveness of outbreak control is reduced.
\end{abstract}

\section{Introduction} %%%%%%%%%%%%%%%%%%%%%%%%%%%%%%%%%%%%%%%%%%%%%%%%%%%%%%%%%%%%%%

How to control infections in schools while allowing pupils as much in-school contact with teachers is an important question that governments throughout the world have grappled with. Balancing the health risks from infection of children in schools with the risks of loss of skills for the young and increase in inequality, the risks to child and parental mental health and the economic and social impact of parents not being able to return to work is a challenging conundrum to resolve~\citep{delve-4}.

Since the start of the pandemic, many countries have incorporated school closures as part of their non-pharmaceutical interventions (NPI) implemented to control disease transmission~\citep{oxford-npi}. A report summarising evidence on schools and transmission from the Children’s Task and Finish Group submitted December 17 to SAGE stated that accumulating evidence was consistent with increased transmission occurring among school children when schools are open, particularly in children of secondary school age; besides multiple data sources showing a reduction in transmission in children following schools' closure for half term \citep{SIS_1_2020,childrens_task_and_finish_group_childrens_2020}.

In England, following the end of the first lockdown, schools fully reopened in September and remained open throughout the autumn term. 
But, in view of the increasing circulation of the Variant of Concern (VOC) B.1.1.7, SAGE told government on December 22 2020 that it is highly unlikely that the stringency of and adherence to the set of NPI measures which were in place from November in England,
which did not include school closures, 
would be sufficient to maintain the effective reproduction number R below 1~\citep{sage-12-2020}.
In early January, in view of the increased transmission of the VOC, the UK government took the decision to postpone an announced programme of testing in schools, which relied in part on rolling out rapid tests using lateral flow devices (LFD) and to close schools till further notice~\citep{nhs_2020}. 
During this period, there was intense discussion about which infection-control policies combining rapid testing and isolation would be both beneficial and feasible to implement in schools~\citep{wise_covid-19_2020,deeks_2021} and how to evaluate their effectiveness, including by randomisation~\citep{bird_2005}. 

It is difficult to disentangle the part played by within-school child-to-child transmission from the knock-on effect of  adult-to-child transmission-chains and increased social contact when schools are opened. A recent modelling study using social contact matrices from surveys at times when schools were opened or were closed suggests that altogether-school-opening could increase the effective reproduction number from $0.8$ to between $1.0$ and $1.5$~\citep{munday_estimating_2021}.

Our work focuses on  within-school transmission and directly addresses the important public health question of how to keep schools open and covid-safe following their reopening on March~8. 
We compare a set of NPI policies that take inspiration from control measures and use of rapid tests that are currently implemented or are being debated and do so with respect to the dual goal of outbreak control and school days lost.
To this end, we propose a realistic agent-based model tailored to the school setting. 
We primarily focus on the bubble-based contact pattern recommended for primary schools in the UK but also 
consider a scenario where bubbles are not feasible to implement.
This latter case is particularly relevant to secondary schools or settings where classrooms are too small to implement effective between-bubble isolation.
Following concerns about compliance with LFD testing,
we also explore a scenario where non-compliance with asymptomatic LFD testing is modelled explicitly.

Our approach differs from the approach taken in \citet{Leng2021} mainly in three aspects. 
Firstly, we model both the test sensitivity and the probability to infect others as functions of the underlying 
viral load of each individual (here pupil) instead of considering these characteristics as independent functions of time since infection. 
This approach allows a realistic correlation between infectivity and test sensitivity driven by the underlying biology.
Secondly, we focus on a primary school setting with a fine-grained population structure. 
We consider policies that act on the level of classes or subgroups of close contacts within classes instead of entire age groups.
Thirdly, the focus of our policy evaluation is on the additional benefit that LFD testing can provide while maintaining the principle of the Test and Trace 
symptom-based isolation instead of substituting it as is considered in \citet{Leng2021}.

By its flexibility, the open access agent-based simulation prototype that we have built will extend to a variety of school and small population environments but here we focus on:
\begin{enumerate}[label={\roman*)}]
    \item setting out the framework of our school SARS-CoV-2 agent-based model, which adapts the viral load based model of~\citet{larremore_test_2021} to small-scale school settings; 
\item a range of testing policies including, as reference, the symptomatic Test~\&~Trace recommendations as well as policies making use of rapid lateral flow tests in combination with specific isolation recommendations; 
\item uncovering the influence of key parameters like infectivity and test sensitivity on the effectiveness of the policies in schools and carrying out an extensive sensitivity analysis to assess the robustness of our conclusions;
\item 
demonstrating that our tailored agent-based modelling allows relative ranking of policies with regards to  offering a good compromise between maintaining infection control and avoiding large number of school days lost, thus providing  inputs to help designing control measures that are more likely to be good candidates for being evaluated in-context by specifically designed studies.
\end{enumerate}

\subsection{Key Assumptions and Policy Context}

We have constructed our model around the following key assumptions, using current literature on SARS-CoV-2 infection and imperfect knowledge on school policy as both are evolving at pace.
\begin{enumerate}%[label={\roman*)}]
\item The proportion of asymptomatic SARS-CoV-2 infections is believed to be higher in children than
in adults \citep{hippich2020,he_2021,wald_2020}.
\item Transmission can occur from asymptomatic infections, both pre-symptomatic and never-symptomatic \citep{arons_2020, sutton_2020, oran_2020}.
\item Transmissibility is related to viral load (VL) \citep{he_2020}. 
\item Transmissibility from symptomatic infections i higher since the VL clearance period is prolonged as compared to asymptomatic cases but peak VL load is the same between both groups~\citep{larremore_test_2021}.
\item Delay from swab-date to PCR-result-date is seldom less than 24 hours \citep{fraser_2021, larremore_test_2021}.
\item Lateral flow devices give a non-quantitative test-result within 30 minutes and are billed as answering a different question than PCR-testing, namely: is a person likely to be contagious~\citep{mina_2021}?
\item Innova LFD tests have been used for screening purposes in nursing homes (now only in conjunction with PCR testing), work-places and primary schools \citep{workplacetesting_2020,department_for_education_2021b, carehome_LDF_2020,dhsc2021}. 
\item Initially, PCR-confirmation of any LFD-positive results was intended. This is no longer the case for on-site LFD tests but LFD positive results obtained through home testing still need to be verified with a follow-up PCR test~\citep[p. 30]{department_for_education_2021b}
\item Plans were well advanced to evaluate (via cluster randomised trial), as the alternative to 10 days of self-isolation at home, that secondary school-pupils who are a close contact of a confirmed case may remain at school provided that their daily LFD tests are negative~\citep{department_education_2021}.
\item The above policy initiative, known as daily-contact-testing, was expected to be trialled in secondary schools which already implement weekly-LFD tests for all pupils, but these plans may be overtaken by a newly
reported policy shift for LFD tests to be used at home twice weekly for
secondary school pupils~\citep{department_for_education_2021b}.

\end{enumerate}

\section{Methods} %%%%%%%%%%%%%%%%%%%%%%%%%%%%%%%%%%%%%%%%%%%%%%%%%%%%%%%%%%%%%%%%%%%

To assess the impact of various policies on the level of individual schools
we adopt an agent-based approach where agents correspond to pupils.
Contacts involving staff are not modelled explicitly for simplicity, as the policy choice is focused on the pupils. 
The overall model is composed of independent sub-models for 
\begin{enumerate}[label={\roman*)}]
\item the contact structure between individual pupils, 
\item viral load and symptom status trajectories during an acute SARS-CoV-2 infection, 
\item the infection probability depending on the latent viral load, 
\item and the sensitivity of the tests (PCR or LFD) that might be required for a policy. 
\end{enumerate}
The time resolution of the overall model is daily, 
i.e.~daily symptom status and viral load are determined at 07:30AM. 
We further assume that any policy intervention (screening tests, isolation) is executed 
before individuals have a chance to meet.
This is an optimistic assumption but justifiable since a recent announcement by the Department for Education includes the possibility of screening tests being sent home from the 15th of March \citep{whittaker_2021}.
We consider a time horizon of 6~weeks which roughly corresponds to the length
of a half-term.

\subsection{Population Model}
\label{sec:population}

The average size of a primary school in England was 281 pupils with an 
average class size of 27 \citep[academic year 2019/20]{govuk_schools_2021}.
The English primary school education consists of six years. 
A typical primary school thus offers either one or two classes per year-group.
We consider a school with two classes per year-group (12 overall)
and 27 pupils per class, i.e.~324 pupils overall.
We further assume that each class is subdivided into 3 \emph{bubbles} 
of 9 pupils each.
Here the term bubble refers to a group of pupils that is isolated as best as
possible from other members of the same class or school~\citep{department_for_education_restricting_2021}.
Although contact tracing is an effective tool to control an epidemic~\citep{ferretti2020},
social distancing and contract tracing within bubbles are deemed unrealistic 
for younger pupils.
The degree of isolation between bubbles depends, among other factors,
on the availability of large enough classrooms and sufficient staff.

\subsubsection{Hierarchical Contact Structure}

We represent the school structure as a three-level hierarchical 
population where each pupil belongs to a bubble nested 
within a class.
The classes, in turn, are nested within a school.
For each of these groups we assume a fixed probability of a risk-contact between 
any pair of members per school day.
\begin{description}[align=left]
    \item[\textbf{Within-Bubble Contacts:}] 
        The highest intensity contact at the bubble level is treated as reference and
we set the daily probability of a risk-contact at the bubble level to $p_{\operatorname{bubble}}=100\%$.
This means that each pair of pupils within a bubble is guaranteed to meet on every single
school day unless a pupil is isolated.  
    \item[\textbf{Within-Class Contacts:}] 
        Each pair of pupils within a class has a daily 
probability of an additional risk-contact of $p_{\operatorname{class}}$.
    \item[\textbf{Within-School Contacts:}]
        Each pair of pupils within the school has a daily 
probability of an additional risk-contact of $p_{\operatorname{school}}$.
\end{description}
The magnitude of the parameters $p_{\operatorname{class}}$ and $p_{\operatorname{school}}$ 
in relation to the $100\%$ chance of having a risk-contact on the bubble level thus jointly 
represent the respective degree of isolation between groups on the different levels 
of the hierarchy.
The contact probabilities on the class- or school level also account for factors 
not explicitly modelled, such as indirect interactions via staff or contacts on the way to or 
from school.
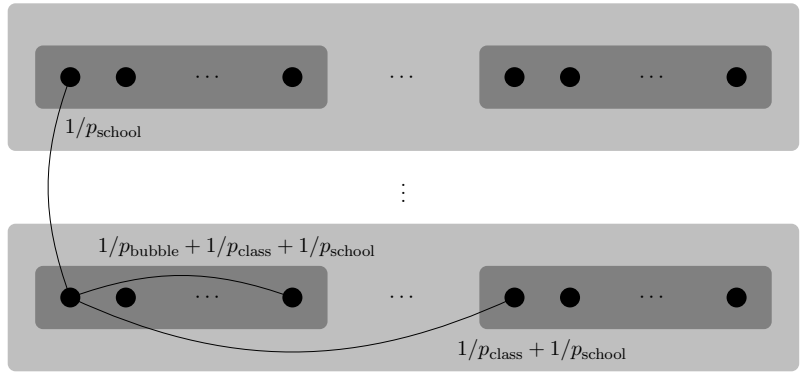
\begin{figure}[h]
    \centering
    \begin{adjustbox}{width=.66\textwidth}
    \begin{tikzpicture}

\tikzset{
    individual/.style={
        circle, fill=black
    },
    bubble/.style={
        rectangle,
        rounded corners,
        fill=gray, thick,
        minimum height=3em,
        text centered,
        text width=4.5cm,
        anchor=center,
        inner sep=1em
    },
    class/.style={
        rectangle,
        rounded corners,
        fill=lightgray, thick, 
        minimum height=7em,
        text centered,
        text width=14cm,
        anchor=center
    }
}

    \node[class,xshift=4cm] (class1) {};
    
    \node[bubble] (bubble1) {};
    \node[individual,xshift=-2cm] (individual11) {};
    \node[individual,xshift=-1cm] (individual12) {};
    \node[xshift=.5cm] (bubble1dots) {$\cdots$};
    \node[individual,xshift=2cm] (individual13) {};
    
    \node[xshift=4cm] (bubble12dots) {$\cdots$};
    \node[bubble,xshift=8cm] (bubble2) {};
    \node[individual,xshift=6cm] (individual21) {};
    \node[individual,xshift=7cm] (individual22) {};
    \node[xshift=8.5cm] (bubble2dots) {$\cdots$};
    \node[individual,xshift=10cm] (individual23) {};
    
    \node[yshift=2cm,xshift=4cm] (vdots) {$\vdots$};
    
    \node[class,yshift=4cm,xshift=4cm] (class2) {};
    
    \node[bubble,yshift=4cm] (bubble3) {};
    \node[individual,yshift=4cm,xshift=-2cm] (individual31) {};
    \node[individual,yshift=4cm,xshift=-1cm] (individual32) {};
    \node[yshift=4cm,xshift=.5cm] (bubble3dots) {$\cdots$};
    \node[individual,yshift=4cm,xshift=2cm] (individual33) {};
    
    \node[yshift=4cm,xshift=4cm] (bubble34dots) {$\cdots$};
    \node[bubble,yshift=4cm,xshift=8cm] (bubble4) {};
    \node[individual,yshift=4cm,xshift=6cm] (individual41) {};
    \node[individual,yshift=4cm,xshift=7cm] (individual42) {};
    \node[yshift=4cm,xshift=8.5cm] (bubble4dots) {$\cdots$};
    \node[individual,yshift=4cm,xshift=10cm] (individual43) {};
    
    \path 
    (individual11.center) edge[bend left=20] node[above,xshift=1cm, yshift=.5em] {$1/p_{\operatorname{bubble}} + 1/p_{\operatorname{class}} + 1/p_{\operatorname{school}}$} (individual13.center)
    (individual11.center) edge[bend right=25] node[above,xshift=4.5cm,yshift=-.75em] {$1/p_{\operatorname{class}} + 1/p_{\operatorname{school}}$} (individual21.center)
    (individual11.center) edge[bend left=20] node[above,xshift=1cm,yshift=0.75cm] {$1/p_{\operatorname{school}}$} (individual31.center);
    
\end{tikzpicture}
    \end{adjustbox}
    \caption{%
        Diagram of the contact structure between pupils; 
        big black dots represent individual pupils; rounded rectangles represent bubbles (dark gray) or
        classes (light gray);
        one representative connection on the bubble-, class-, or school level is drawn as curved line
        annotated with its respective number of daily expected risk contacts.
    }
    \label{fig:school-structure}
\end{figure}

\subsubsection{Parameter Choice}

To the authors' knowledge, data on the number of per-class or per-school contacts of 
young children are not available and would highly depend on context-specific definition of 
what is assumed a `risk-contact'. Parameter choices thus have to remain somewhat arbitrary.
For our primary analysis, we chose $p_{\operatorname{class}} = 3/(|\operatorname{class}| - 1)$ and $p_{\operatorname{school}} = 1/ (|\operatorname{school}| - 1)$.
This implies that each pupil has an expected daily number of $3$ additional daily risk-contacts
within their class and one additional risk contact with any pupil in the school 
($8 + 3 + 1 = 12$ in total).
The expected number of contacts decreases naturally as pupils start to go into isolation 
(see Section~\ref{sec:policies}).
The adjacency matrix of the school structure used for the primary analysis is 
shown in Figure~\ref{fig:adjacency-matrix}.
We also investigate a scenario where effective between bubble isolation is impossible and the whole class becomes one bubble
(see Section~\ref{sec:sensitivity-one-bubble}).

\subsection{A Model for Viral Load and Symptoms}
\label{sec:vl-model}

\newcommand{\vl}{\operatorname{\scalebox{.8}{VL}}}
\newcommand{\logten}{\ensuremath{\log_{10}}}
\newcommand{\lli}{ \ensuremath{\operatorname{LLI}} }

Data on the evolution of viral load ($\vl$) in children during an acute
infection with SARS-CoV-2 are rare but cross-sectional data suggest that there is no 
substantial difference between VL of symptomatic children and adults~\citep{baggio_sars-cov-2_2020,jones_analysis_2020}. 
We thus build on available evidence for $\vl$-trajectories over time in
adults and the model proposed in \citet{larremore_test_2021}.
Here, each individual's $\vl$-trajectory is determined by a set of pivot points 
with ordinates on the $\logten(\vl)$ scale
and subsequent linear interpolation of the pivot points.
The pivot points are
\begin{description}[align=right,labelwidth=\widthof{\textbf{start of fast exponential growth:}}]
    \item[\textbf{start of fast exponential growth:}] $\big(t_1, \logten(\vl_{\,\operatorname{start~fast~growth}})\big)$ 
    \item[\textbf{peak log-10 VL:}] $\big(t_2, \logten(\vl_{\,\operatorname{peak}})\big)$
    \item[\textbf{clearance point}:] $\big(t_3, \logten(\,\lli)\big)$
\end{description}
where $\lli$ is the viral load at the \emph{lower limit of infectivity},
a point were the infection probability is zero or close to zero 
(see Section~\ref{sec:infectivity-submodel}).
\citet{larremore_test_2021} used $\lli=10^6$ and $\vl_{\,\operatorname{start~fast~growth}}=10^3$.
We assess the sensitivity with respect to $\lli$ in Section~\ref{sec:sensitivity-lli}.
The distribution of the $\logten(\vl)$-trajectories is given implicitly
by the following sampling procedure.

Firstly, it is determined whether the trajectory will ultimately become 
symptomatic by sampling from a Bernoulli distribution with a probability 
$p_{\operatorname{symptomatic}}$.
Secondly, the first pivot time $t_1$ is sampled uniformly between 
$2.5$ and $3.5$ days after the infection time $t_0=7.5/24$.
Here we deviate from \citet{larremore_test_2021} since they
consider a continuous-time model while we discretize all relevant values 
at 07:30AM.
Thirdly, peak $\vl$-delay with respect to $t_1$ is sampled as
$t_2 - t_1 = 0.5 + \min(3, X)$ where $X\sim \operatorname{Gamma}(1.5)$.
The corresponding peak log-10 viral load, $\logten(\vl_{\operatorname{peak}})$, 
is sampled uniformly between $[7, 11]$.
The timing of the third pivot $t_3$ is then sampled conditional on whether or 
not an individual is symptomatic:
For asymptomatic cases, $t_3 - t_2\sim\operatorname{Unif}(4, 9)$. 
For symptomatic cases, a symptom onset time with delay 
${t_{\operatorname{symptoms}}-t_2\sim \operatorname{Unif}(0, 3)}$ 
is sampled to determine the time to symptom onset and this symptom onset delay 
is added to $t_3$.
The latter implies that symptomatic cases have a slower clearance of their 
peak~$\vl$~but the same peak~$\vl$.
For symptomatic individuals, we assume that the symptomatic period lasts 
from the sampled onset time until the viral load drops under $\lli$.

\begin{figure}[h]
    \centering
    \includegraphics[width=\textwidth]{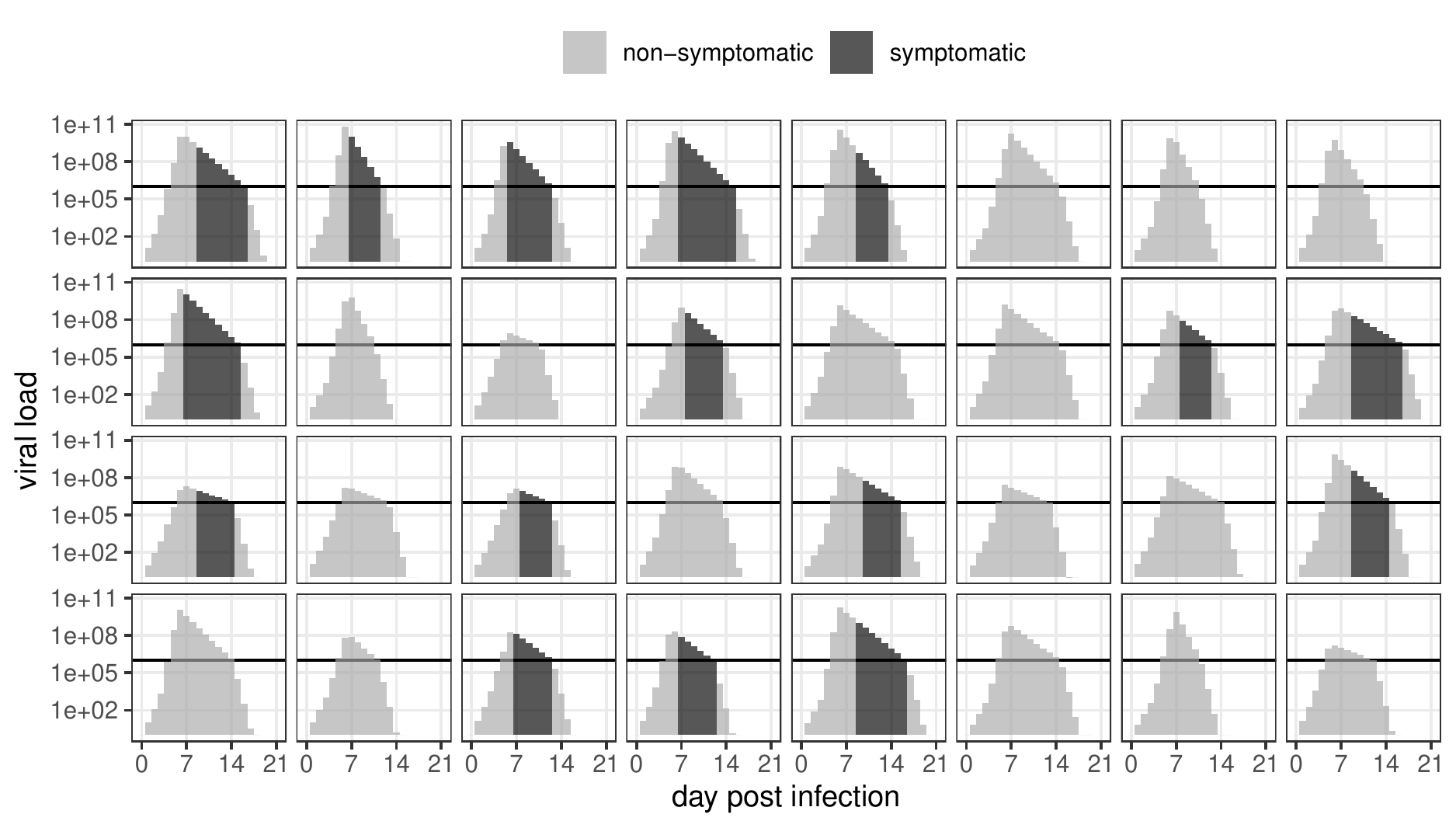}
    \caption{%
        32 randomly sampled $\vl$-trajectories under the Larremore~model 
        assuming that $50\%$ of children are asymptomatic ($p_{\operatorname{symptomatic}} = 0.5$);
        horizontal solid black line corresponds to $\protect\lli=10^6$. For the symptomatic trajectories, the VLs from the simulated day of the symptom onset till the day when the VL reaches below the LLI are highlighted as vertical black lines
    }
    \label{fig:sample-trajectories}
\end{figure}

We set the initial $\vl_{7.5/24}=1$ and 
assume that that $\logten(\vl)$ drops linearly to $0$ within 3~days after 
reaching the clearance point $t_3$ (not specified in~\citet{larremore_test_2021}).
Outside of this interval, $\vl_t=0$, i.e.~$\logten(\vl_t)=-\infty$
(see Figure~\ref{fig:sample-trajectories} for example trajectories).
We assume a daily rate of $1\%$ for Covid-like symptoms like dry cough etc.~due 
to non-Covid-related causes. 
As a sensitivity analysis, we also consider a case where additional variability is superimposed on the VL
trajectories to create heavier tails of the VL distribution (see Appendix~\ref{apx:sensitivity-heavy-tails}).

\subsection{Infectivity Model}
\label{sec:infectivity-submodel}

Given the short time-horizon of only 6 weeks, we assume that individuals who 
already went though an infection are no longer susceptible to infection 
(`short term immunity').
We model the probability to infect a susceptible individual during a risk-contact
(`infection probability') as function $f(\vl_t)$ of the infected individual's
latent viral load on the day of the risk-contact $t$.
\citet{larremore_test_2021} conduct sensitivity analyses for different functional 
forms of $f$ and base their main results on a model where the infection 
probability is assumed to be proportional to $\logten(\vl_t)$ if a lower limit of infectivity, 
$\lli$, is exceeded, i.e.,
\begin{align}
    f_{\operatorname{Larremore}}(\vl_t) 
        := \min\big(1, \max\big(0, \gamma \, \big(\logten(\vl_t) - \logten(\lli) \big)\, \big)\, \big) \ .
\end{align}
Whenever the $\lli$ is fixed externally, 
infectivity only depends on the choice of~$\gamma$, referred to henceforth as infectivity parameter.
We follow the suggestion of \citet{larremore_test_2021} to match~$\gamma$ to a target school-level reproduction number~$\rzero$ (see Section~\ref{sec:calibration-infectivity}).
Here, the reproduction number is defined as the average number of infections from a 
given index case in a completely susceptible school population, i.e.~no isolation or immunity, followed for 21 days.

\iffalse % not inlcuded
\marginnote{%
    do we want to keep this, \citet{peto_2021} seem to rather support the piece-wise linear model
}
Another candidate for the function form is a logistic regression 
\begin{align}
    f_{\operatorname{log-reg}}(\vl) 
        := \operatorname{logit}^{-1}\big(\gamma \, \logten(\vl) + c_{\operatorname{infectvity}}\big) \ .
\end{align}
The proposed logistic curve varies smoothly between $0$ and $1$ but never attains these boundary values.
We thus suggest to link the intercept to $\lli$ by imposing the constraint
$f_{\operatorname{log-reg}}(\lli) = 1/1000$, i.e., $c_{\operatorname{infectivity}} = \log(0.001/0.999) - \gamma\,\logten(\lli)$.

\fi % not included

\subsection{A Model for the Screening Test Characteristics}
\label{sec:sensitivity-submodel}

Sensitivity of LFD tests has been shown to depend on viral load
\citep{liverpool_covid-19_2020,peto_2021}.
This is a crucial feature since a joint dependence of 
test sensitivity and infection probability on the latent viral load trajectories implies a positive correlation between the two. Following data presented in~\citet{peto_2021}, we consider a 
logistic regression model for the functional from $g(\vl)$ of 
the test sensitivity as function of viral load
\newcommand{\lod}{\operatorname{LOD}}
\begin{align}
    g(\vl_t) 
        := \operatorname{logit}^{-1}\big( 
	        \beta_{\operatorname{test}} \, \logten(\vl_t) + c_{\operatorname{test}}
        \big) 
    \label{eq:sensitivity-submodel}
\end{align}
where $\beta_{\operatorname{test}}$ is the VL slope on the $\logten$ scale and $c_{\operatorname{test}}$
the intercept.
We calibrate the sensitivity curve by fitting it to 
cross-sectional data assuming that $50\%$ of individuals are asymptomatic
(see Section~\ref{sec:calibration-sensitivity}).
The specificity of LFD tests can generally be considered fairly high
and we assume a fixed value of $0.998$~\citep{liverpool_covid-19_2020}.

Concerns have been raised that, due to person specific effects, assuming independence between results of repeated tests is unrealistic
(see comments by Jon Deeks, et al.~on \citet{kmietowicz2021a}).  
In our model, there is an implied dependence between subsequent tests results of an individual as these are functionally linked to the latent VL. 
Importantly, within-individual autocorrelation of test results will directly
affect the performance of policies which rely on repeated screening tests:
if the autocorrelation is high, repeated testing of the same individual has
less benefit than under a model with less autocorrelation because even a 
screening test with low sensitivity might be able to identify pre-symptomatic
infections after two or three days of daily testing. 

\begin{figure}[h]
    \centering
    \includegraphics[width=\textwidth]{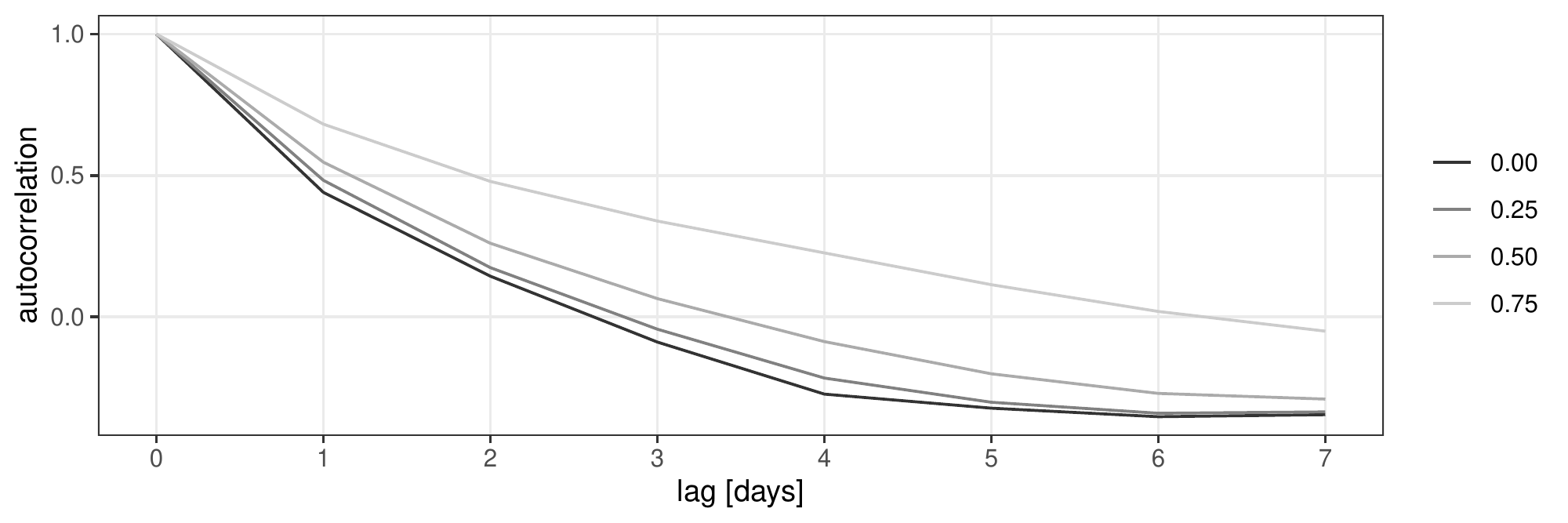}
    \caption{%
        Simulated autocorrelation functions for varying $a$ under 
        $50\%$ asymptomatic cases and the Larremore-model 
        (only relevant pre-symptomatic phase);
        assumed mean sensitivity of the LFD test used is $60\%$ 
        (see Section~\ref{sec:calibration-sensitivity});
        individuals are infected on Mondays and tested daily 
        (except weekends); maximal 21 days follow-up.
    }
    \label{fig:test-autocorrelation}
\end{figure}

We explore the impact of increased within-subject autocorrelation of test results by imposing an auto-regressive structure on the screening test sensitivity. For each individual and each time point $t$, we first look back if there has been a LFD test done within a time-window consisting of the three days previous to $t$. If no testing took place in the window,  equation~\eqref{eq:sensitivity-submodel} is not modified. 
If one or several tests were carried out in that window, we amend equation~\eqref{eq:sensitivity-submodel} as follows:
let $x^{i}_t$ be the most recent LFD test result in the time-window for individual $i$ 
 ($x^i_t=0$ for negative, $x^i_t=1$ for positive). 
We then define 
\begin{align}
    \widetilde{g}\big(\vl_t, x_t^i\big) 
        := \begin{cases}
            g(\vl_t), \quad \text{if no test done in time-window} \\
            (1 - a)\,g(\vl_t) + a\, x_t^i\qquad \text{else}
        \end{cases} \ . 
    \label{eq:sensitivity-submodel-ar}
\end{align}
Here $a, \; 0 \leq a \leq 1$ is the auto-regression coefficient and a large $a$ implies that the
results of repeated tests are heavily biased towards the respective last result.
The effect of $a$ on the autocorrelation of repeated test results is visualized
in Figure~\ref{fig:test-autocorrelation}.
Note that even for $a=0$ the smoothness of the $\vl$-trajectories implies implicit substantial 
autocorrelation between repeated tests.
If a testing scheme only re-tests the same individual  after a time gap between
individual tests larger than 3 days, the test characteristics remain unchanged.
In particular, cross-sectional testing of a population 
(as done with the Liverpool study) is not affected.
Testing policies that rely on repeated testing of individuals within the
specified time-window are, however, affected since the
chance of repeated false negative  findings is increased when the initial test was itself a false
negative. 
This is particularly important when considering policies like test for release 
(see Section~\ref{sec:test-and-release}).

\subsection{Policies}
\label{sec:policies}

We compare different test and isolation policies that have been discussed
in the context of reopening primary schools in England.
For simplicity we do not consider multi-level strategies with
policies on the class or school level but only policies that intervene
on the bubble level.

In all cases, we assume that the swab for a confirmatory PCR follow-up test
is taken on the day of symptom onset or of testing positive with a LFD screening test \citep{gov_covid_test_2021}. 
Note that a PCR follow-up test is no longer required for on-site LFD testing according to the latest guidance released by the~\citet{department_for_education_2021b}.
We assume a turnaround time for PCR tests of two days (including the swab-day) \citep{fraser_2021}. 
The isolation time for PCR-confirmed cases is 10 days starting with the PCR-swab-day which we assume to be the same as the LFT-swab-day~\citep{nhs_2020}.

PCR tests are more sensitive than antigen-based 
screening tests and we assume a flat sensitivity of $97.5\%$ above a limit of detection of $300 \operatorname{cp}/\operatorname{ml}$ and a specificity of $100\%$ (see e.g.~\citet{fda_2020} for a detailed listing of different assays' limit of detection).
Across all policies we assume that any pupil who becomes symptomatic 
is immediately isolated at home before school on the day of symptom onset and
a swab for a follow-up PCR test is taken.
Such a pupil only returns to school after isolating for either 10 days from their swab date (positive result)
or 2 days (negative swab test, only isolated during the PCR turnaround time). 

\subsubsection{Reference Policy}
\label{sec:reference-policy}

The reference policy follows the current Test~\&~Trace recommendations. 
Its implementation assumes that the close contacts of an index case are the 8 other children in the bubble of the index case. This reference policy does not use LFD tests and solely relies on symptom-driven isolation.
If an index case shows symptoms and starts their self-isolation period,
the remaining members of the bubble (and class) continue to attend school until the test
result of the symptomatic index case becomes available.
Only if the index case's PCR test turns out to be positive do the remaining individuals
in the bubble isolate for the remaining 8 days. 
Newly symptomatic cases while in isolation are also checked with PCR tests and newly
emerging PCR-positive results reset the isolation clock for the entire bubble. 

\subsubsection{Extended Weekend}
\label{sec:extended-weekend-policy}

As a simple-to-implement variant of the reference policy,
we consider an extension where the entire school is closed on Thursdays and Fridays, and teaching switched to online.
Otherwise the same procedures as under the reference policy apply.
This effectively introduces a mini-lockdown of four days over the extended weekend
which facilitates the identification of symptomatic cases before they can
spread the virus in school.

\subsubsection{Mondays Screening}
\label{sec:monday-screening-policy}

To assess the added benefit of regular screening tests we consider the
reference policy extended by regular rapid LFD screening tests on Mondays before going into class
for every pupil in the school (except those already isolating).
Since LFD tests are considerably more specific than mere symptoms, we assume that a positive
LFD test result for an index case leads to an immediate isolation and return home of the entire
bubble of the index case.
The bubble (and the index case)  return to school either after 2 days if the index case's PCR
test turns out to be negative (2 days isolation) or after the full 10 days of isolation
if the index case's PCR test turns out to be positive.
Note that due to the 7 days gap between the screenings, 
this policy would not be affected by the introduction of additional retest autocorrelation 
(see Section~\ref{sec:sensitivity-submodel}).

\subsubsection{Mondays and Wednesdays Screening}
\label{sec:monday-thursday-screening-policy}

Policies with multiple screening tests per week have been discussed. 
Austria, for instance, has laid out a plan for twice-weekly screening tests at schools \citep{haseltine_2021}.
We thus also consider a policy that extends the reference policy by twice-weekly 
testing on Mondays and Wednesdays.
In this case, the results of the Wednesday screening will be affected if we include positive autocorrelation
(${a >0}$) between
the tests (see Section~\ref{sec:sensitivity-submodel}).

\subsubsection{Test for Release}
\label{sec:test-and-release}

Finally, we consider a policy that we refer to as `test for release'.
Such an approach was proposed in early 2021
to avoid preemptive bubble isolation in schools~\citep{department_education_2021}.
Test and release avoids bubble-isolation completely.
Instead, under a test for release policy
members of the bubble around symptomatic or  LFD-positive index cases are followed up using daily LFD testing.
No preemptive isolation on the bubble level is imposed.
Only newly symptomatic or LFD-positive individuals isolate, while the
remainder of the bubble attends school.
Symptomatic LFD-positive cases are told to self-isolate immediately and
are then followed up with PCR tests as under the default strategy. 
The bubble-wide LFD testing starts on the day of the index case's 
triggering event (either symptom onset or a positive LFD test) and 
continues for up to 7 school days, 
i.e.~neither Saturdays nor Sundays count towards the LFD follow up days.
Daily bubble-contact testing is terminated early if the index case's follow-up PCR test turns out to 
be negative (after 2 days).

\subsection{Model Implementation}

We implemented the individual components of the overall model in a package~\citep{kunzmann2021a} for the programming language \texttt{Julia}~\citep{Julia2017}.

For each scenario, we reran the simulation 250 times to capture the variability of the outcome measures of interest. 
Each run was conducted by first initialising the individuals and the school structure according to the specified scenario. 
The start day is $0$ and we assume that no pupils are infected at onset.
For each day of the simulation (6 weeks, 42 days) we then
\begin{enumerate}
    \item Randomly sample new school-external infections for each pupil. We use a fixed Binomial probability for each pupil and day of $1/324/7$ which results in one expected external infection per week. 
    \item If school day (default: Monday to Friday): Execute the test and isolation policy. 
    This entails checking for symptomatic cases and/or conduct LFD testing if specified.
    Isolation of individuals or bubbles is then handled according to the respective policy.
    \item If school day: Randomly sample risk contacts for pupils not isolating according to the school contact structure, i.e., on the bubble level, the class level, and the school level.
\end{enumerate}
The plots used in this manuscript were generated using a combination of \texttt{R}~\citep{r2020} and \texttt{Julia} and the source code is available online~\citep{kunzmann2021b}.

\section{Results} %%%%%%%%%%%%%%%%%%%%%%%%%%%%%%%%%%%%%%%%%%%%%%%%%%%%%%%%%%%%%%%%%%%%

The baseline scenario considered is based on a fraction of $50\%$ asymptomatic cases~\citep{hippich2020},
an expected number of weekly community infections of $1$,
$\lli=10^6$,
and no additional within-subject autocorrelation of test results ($a=0$). We set $\rzero=3$ and fix the mean LFD test sensitivity to be 60\%. Recall that $\rzero$ has been calibrated specifically for our school-based three-level contact pattern and choice of probability of contacts between pupils, as described in 
 Section~\ref{sec:calibration-infectivity}. We then consider extensive sensitivity analyses around this baseline scenario.
 
 \subsection{The Baseline Case}
\begin{figure}[H]
    \centering
    \includegraphics[width=\textwidth]{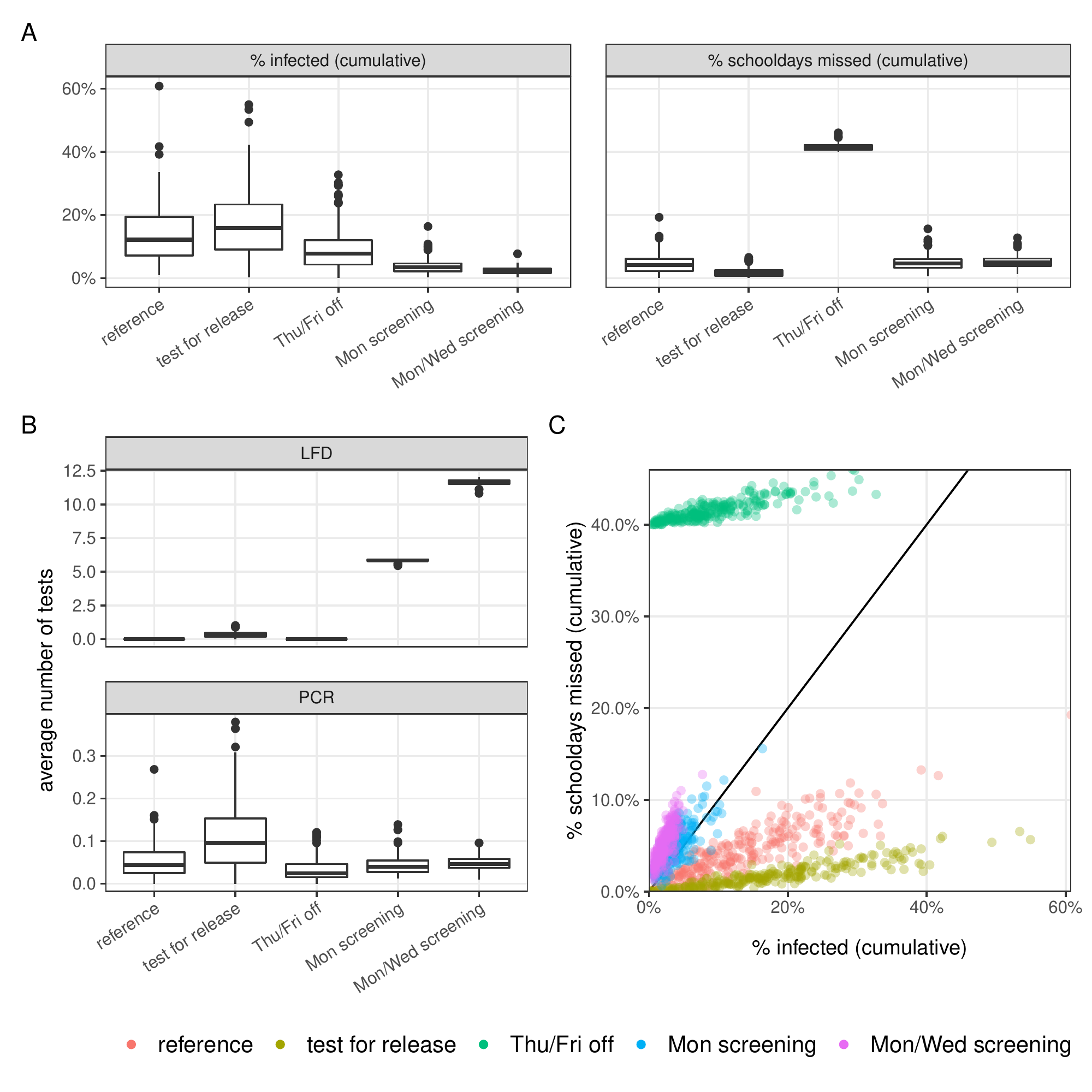}
    \caption{%
        Results for the baseline scenario with $\rzero=3$, mean pre-symptomatic 
        LFD test sensitivity of $0.6$, and $50\%$ asymptomatic cases over the 6-week horizon;
        A: Box-plots of the distribution of the proportion of infected pupils and the
        the proportion of schooldays missed;
        B: Box-plots of the average number of PCR and LFD tests per pupil;
        C: Scatterplot of the proportion of missed schooldays against the proportion of infected pupils of individual simulation runs, black line indicates first bisector.
    }
    \label{fig:results-main}
\end{figure}

We first look at the relative effectiveness of the different policies in terms of
containing the number of infections among pupils and the number of school days lost, the main criteria of interest for comparing policies.
Figure~\ref{fig:results-main} shows box plots of the marginal distribution of the
proportion of pupils infected (top left), 
the proportion of school days lost (top right),
and the average number of LFD and PCR tests per pupil (bottom left and right)
for the 5 policies over the simulated 6~weeks time horizon. 

In terms of containing school outbreaks, we see that both the reference policy and the test for release policies do not fully succeed in containing outbreaks, the reference policy, 
which relies on symptomatic PCR testing and bubble isolation, 
performing slightly better than the LFD-based test for release approach. 
Additional regular weekly asymptomatic testing on Mondays clearly improves outbreak control over the reference policy
with a similar proportion of schooldays missed and a higher LFD test burden per child.
A second regular screening on Wednesday improves containment only marginally while doubling the LFD test burden. 
The extended weekend scenario gives intermediate results in terms of containment 
while increasing considerably the number of school days lost. 

The health impact associated with Covid-19 is largely determined by age and is much 
smaller in young children. 
This implies that  focusing solely on the number of infections over the 6-weeks period is insufficient as a performance measure for policies 
in a primary school context. 
The various policies' trade-off between schooldays missed and 
the effectiveness of the containment of new outbreaks is a key performance indicator.
The fraction of schooldays missed is plotted against the fraction of ultimately 
infected individuals in Figure~\ref{fig:results-main}~C.
The proportion of schooldays missed is positively correlated with the cumulative
number of infections
since all policies incorporate some form of isolation component once new cases are 
detected.
Policies clustering above the first bisector favour containment over attendance. 
Interestingly, the reference policy is dominated by `test for release' when considering
the trade-off between attendance and containment although both fare poorly in terms
of their capability to control new outbreaks in high-infectivity scenarios
Both the reference policy and test for release clearly favour attendance over containment
with test for release being the most extreme.

We have chosen to present the cumulative number of infections.
An alternative metric to evaluate containment would be the mean daily 
number of infectious and non-isolating pupils.
We found that this metric correlates very strongly with the presented cumulative proportion of
infected pupils (data not shown) and a separate discussion is therefore not warranted.

\clearpage
\subsection{Sensitivity Analyses}

\subsubsection{Infectivity, Symptoms, and Test Sensitivity}
\label{sec:sensitivity-r-sympt-sens}

\begin{figure}[H]
    \centering
    \includegraphics[width=\textwidth]{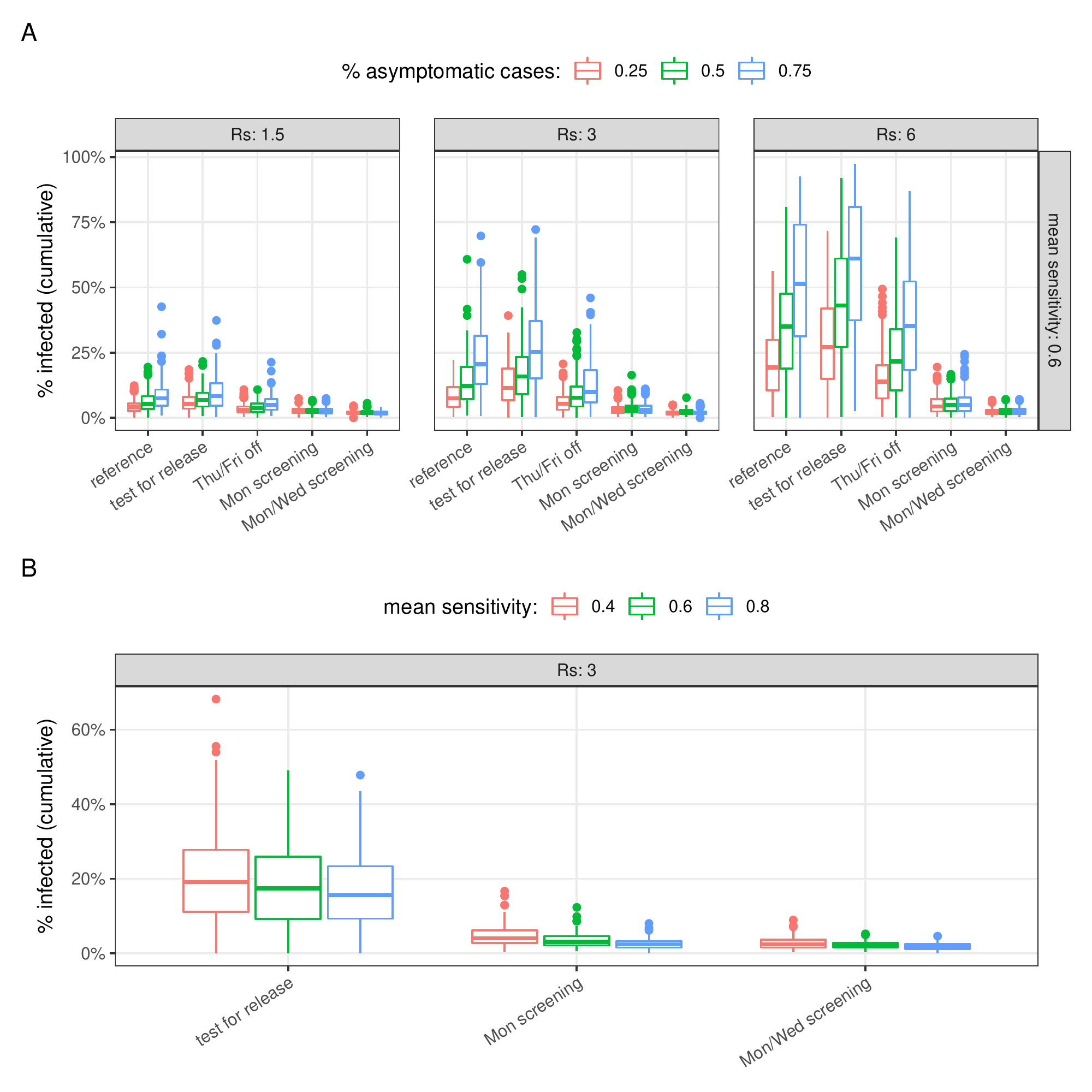}
    \caption{%
        A: Proportion of infected pupils by~$\protect\rzero$ and the fraction of asymptomatic cases 
        over a 6-week horizon for a mean LFD test sensitivity of 60\%;
        B: Proportion of infected pupils by mean LFD test sensitivity for $\protect\rzero=3$ (LFD-based policies only).
    }
    \label{fig:sensitivity-symptomatic-lfd-sensitivity}
\end{figure}

Data on the actual LFD test sensitivity and the fraction of asymptomatic children are scarce 
and evidence on between-student infectivity ($\rzero$) is difficult to map to the particular school structure 
considered here.
We thus investigate the stability of the results with respect to these three key parameters 
over a range of values (see Figure~\ref{fig:sensitivity-symptomatic-lfd-sensitivity}).

The differences between policies with infectivity ($\rzero$) and are most pronounced in 
the highest infectivity scenario ($\rzero=6$) but the relative performance of the different policies
remains stable.

As expected, an increased proportion of asymptomatic cases leads to a deterioration of infection containment for all policies. 
It is particularly interesting that increasing the proportions of asymptomatic cases from 25\% to 75\%
affect both the reference policy and test for release in a similar manner. 

Consistent with the findings of \cite{larremore_test_2021}, Figure~\ref{fig:sensitivity-symptomatic-lfd-sensitivity}~B demonstrates that the actual sensitivity 
of the LFD test employed is secondary to other factors. 
Under test for release this is mainly due to the low number of LFD tests conducted on average over the
time period considered (see Figure~\ref{fig:results-main}~B).
The relative impact is higher in scenarios with regular screening due to the higher number of tests
but the absolute impact is small compared to the between policy differences.

\subsubsection{Re-Test Autocorrelation} %---------------------------------------------------
\label{sec:sensitivity-autocorrelation}

\begin{figure}[H]
    \centering
    \includegraphics[width=\textwidth]{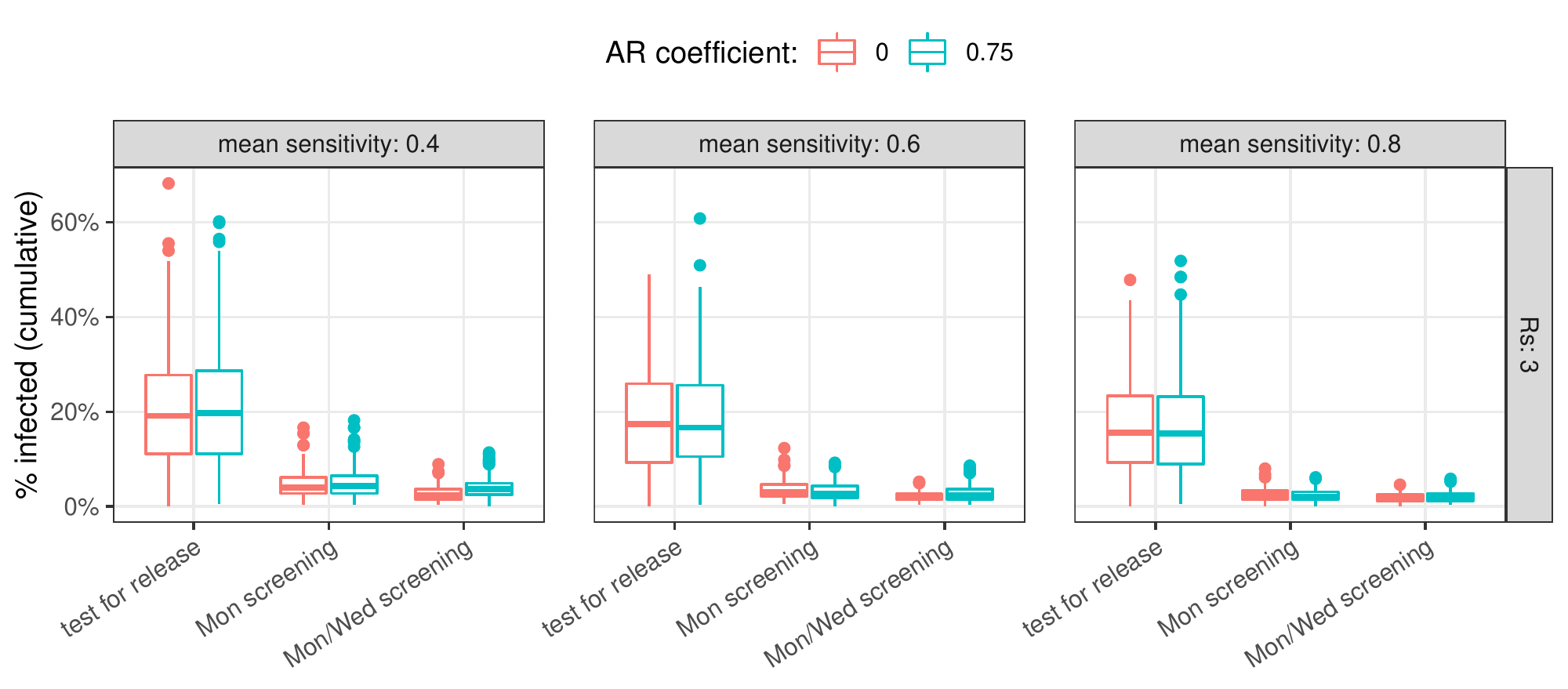}
    \caption{%
        Proportion of infected pupils by autocorrelation coefficient $a$ and mean LFD test sensitivity  
        over a 6-week horizon for a mean LFD test sensitivity of 60\% and $\protect\rzero=3$.
    }
    \label{fig:sensitivity-autocorrelation}
\end{figure}

As discussed in Section~\ref{sec:sensitivity-submodel}, 
concerns have been raised whether the
amount autocorrelation between subsequent LFD tests that is 
implied by the dependency of LFD test sensitivity on VL is sufficient.
Figure~\ref{fig:sensitivity-autocorrelation}~B shows results both 
with strong additional re-test autocorrelation ($a=0.75$) and without ($a=0)$ according to the 
model extension discussed in Section~\ref{sec:sensitivity-submodel}.

A value of $0.75$ for the auto-regressive component is fairly high and 
implies that the probability of a positive test result within 3 days of a negative initial 
result is at most $25\%$ - even if the test characteristics imply a sensitivity of $100\%$.
This relatively extreme scenario was chosen since the intrinsic dependence between 
repeated tests is already high (see Figure~\ref{fig:test-autocorrelation}) 
and smaller values of $a$ have even less impact on results (data not shown).

The level of additional re-test autocorrelation does not affect the results substantially
across the considered values of mean pre-symptomatic LFD test sensitivity.
A difference is only discernible for `test for release' in scenarios with
relatively bad operating characteristics of the LFD test 
(mean sensitivity of $40\%$, see leftmost panel in Figure~\ref{fig:sensitivity-autocorrelation}~B).

\subsubsection{Only One Bubble Per Class} %---------------------------------------------------
\label{sec:sensitivity-one-bubble}

\begin{figure}[H]
    \centering
    \includegraphics[width=\textwidth]{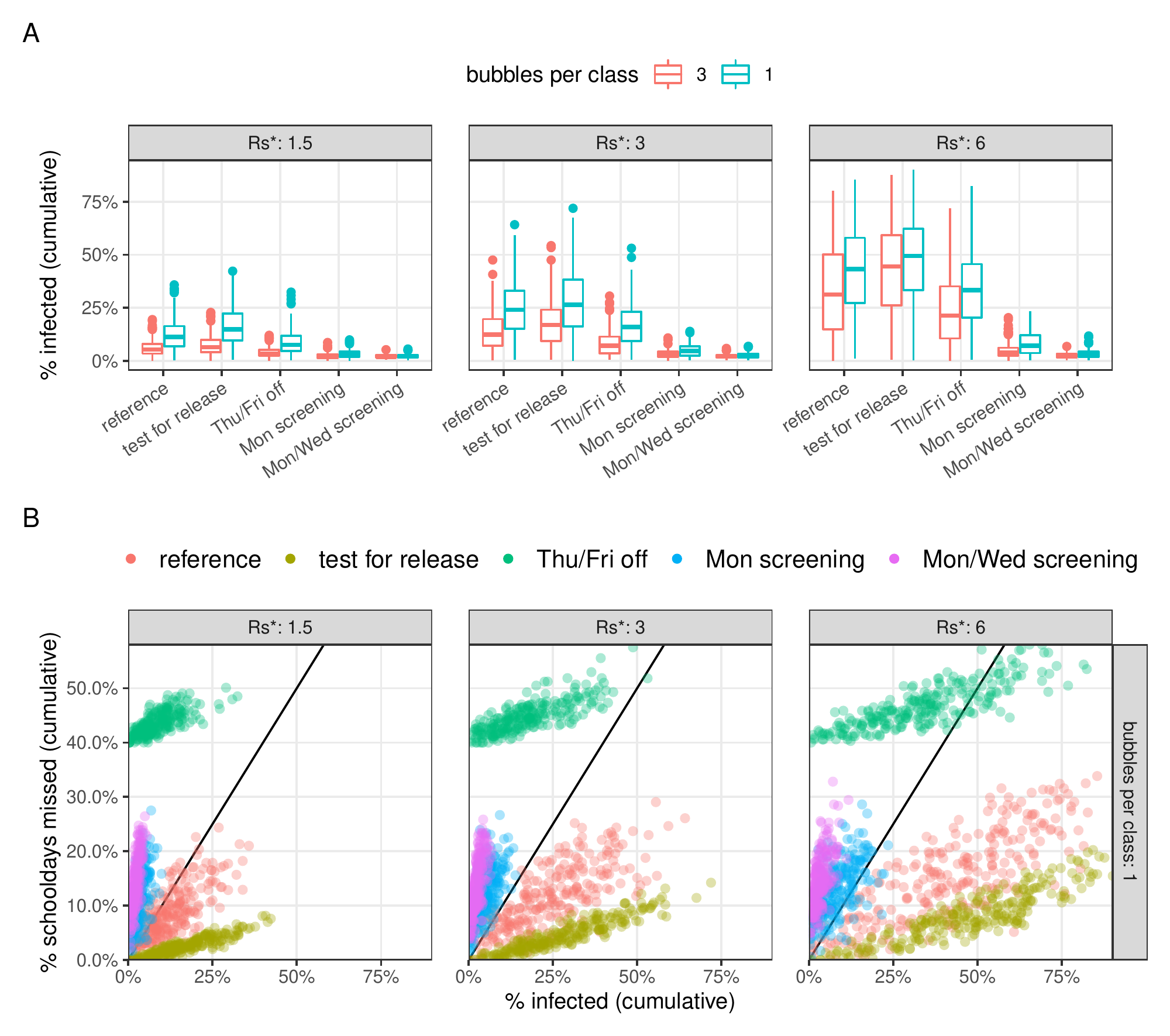}
    \caption{%
        Sensitivity of results with respect to the class structure (one bubble or three per class);
        $\protect\rzero^*$ in the plot labels refers to the $\rzero$ under the three-bubbles per class scenario.
    }
    \label{fig:sensitivity-one-bubble}
\end{figure}

The proposed bubble isolation concept might be infeasible in individuals institutions
for a number of reasons.
In primary schools, there might not be enough room physically to separate groups of young 
children or it might turn out that additional staff is required to enforce effective
separation between bubbles during class.
Moreover, in secondary schools, the concept of `bubble' is not relevant.
Instead \citet{Leng2021} considered policies on the year-group level with up to 200 individuals.
Here, a class-based approach could be a compromise.
We thus consider the case of a single bubble per class. 
This means that each pair of pupils in the class has at least one daily risk contact
and that all policies are executed at the class level.
The altered class structure leads to an increase in expected daily risk-contacts per 
pupil as compared to a class with 3 bubbles of 9 pupils each.
This, in turn, increases the $\rzero$ for any given infectivity constant~$\gamma$.
For the sake of comparability between scenarios, we do not re-calibrate~$\rzero$
to this new `one bubble' class structure.
The results are given in Figure~\ref{fig:sensitivity-one-bubble}.

Our simulations indicate that the increased number of expected daily contacts more than offsets the wider scope of policy execution (i.e. isolation of the whole class if there is a positive case). 
In consequence, the containment properties of most policies are worse than under an
effective bubble partition of the whole class (see Figure~\ref{fig:sensitivity-one-bubble}).
Jointly, the increased number of contacts and the wider scope of the respective 
isolation policies lead to an increased variability of outcomes but the
qualitative results on relative effectiveness of the policies remain unchanged.

\subsubsection{LFD-Test Compliance} %===================
\label{sec:sensitivity-lfd-compliance}

\begin{figure}[H]
    \centering
    \includegraphics[width=\textwidth]{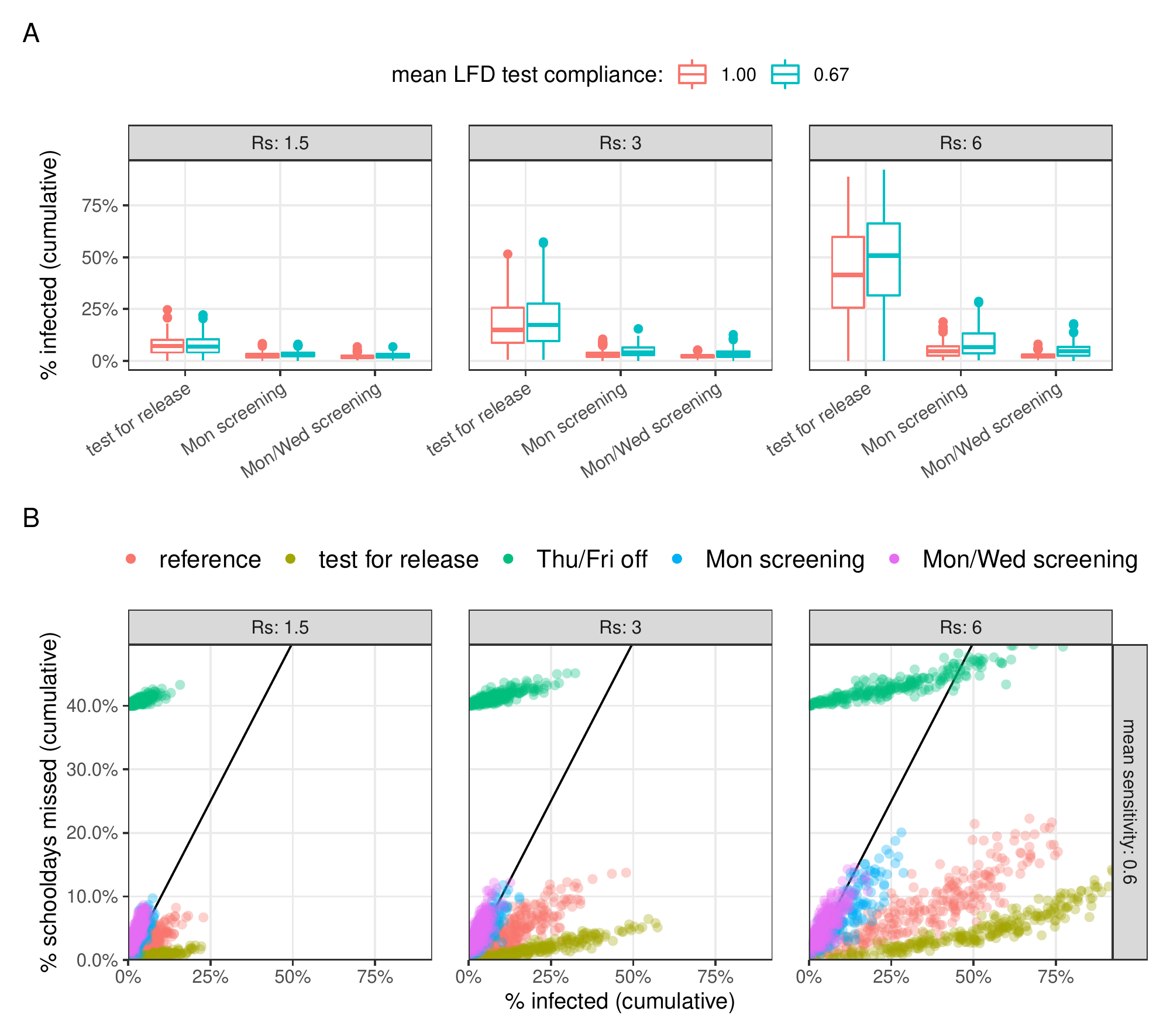}
    \caption{%
        Sensitivity of results with respect to mean LFD test compliance;
        $\protect\rzero$ in the plot labels refers to the $\rzero$ under the three bubbles per class scenario;
        only the LFD-based policies are affected by compliance but we show results for all policies in the scatterplots for easier reference.
    }
    \label{fig:sensitivity-lfd-cmpliance}
\end{figure}

All preceding scenarios assumed perfect compliance of individuals with the respective
testing schemes (both PCR and LFD).
PCR tests are usually conducted as follow-up to either becoming symptomatic
or receiving a positive result from a screening test and it is reasonable to assume a high
compliance rate.
For asymptomatic LFD tests, this is not necessarily the case and compliance rates of children and parents as
low as $40\%$ cannot be ruled out in practice~\citep{wheale_2021}.
We explore the impact of non-compliance by assuming that each pupil has a latent 
`LFD test compliance probability' of actually carrying out a policy-recommended LFD test.
For simplicity, we also assume that failure to comply with a LFD testing request does not affect their compliance with other recommendations such as isolation, 
and that non-compliant children are attending schools along the compliant children, a worst case scenario.
It is reasonable to assume that the willingness to comply with LFD tests varies between pupils
and we model this by drawing individual compliance probabilities from a U-shape dispersed Beta distribution with mean $0.66$ (see Section~\ref{apx:compliance} for results and details of the implementation).
The results are given in Figure~\ref{fig:sensitivity-lfd-cmpliance}

Increased non-compliance reduces the effectiveness of policies slightly without
changing the relative efficiency of different policies.
Notably, the reduced compliance again affects regular screening policies more than
test for release in relative terms since the number of tests scheduled tends to be higher.
This leads to both regular screening policies being clustered on or below the first bisector
(compare Figure~\ref{fig:sensitivity-lfd-cmpliance}~B, $\rzero=3$ and Figure~\ref{fig:results-main}).

\subsubsection{Lower Limit of Infectivity} %===================
\label{sec:sensitivity-lli}

A crucial feature of the overall model is the assumed relation between the test 
sensitivity and the infection probability -
if it can be assumed that a LFD test is highly sensitive while the infection probability
is still small, test-based policies for containment are easier to implement.
We thus also explore a scenario, where the $\lli$ is much lower, $\lli=1000$, 
instead of $\lli=10^6$ as suggested by \citet{larremore_test_2021}.

\begin{figure}[H]
    \centering
    \includegraphics[width=\textwidth]{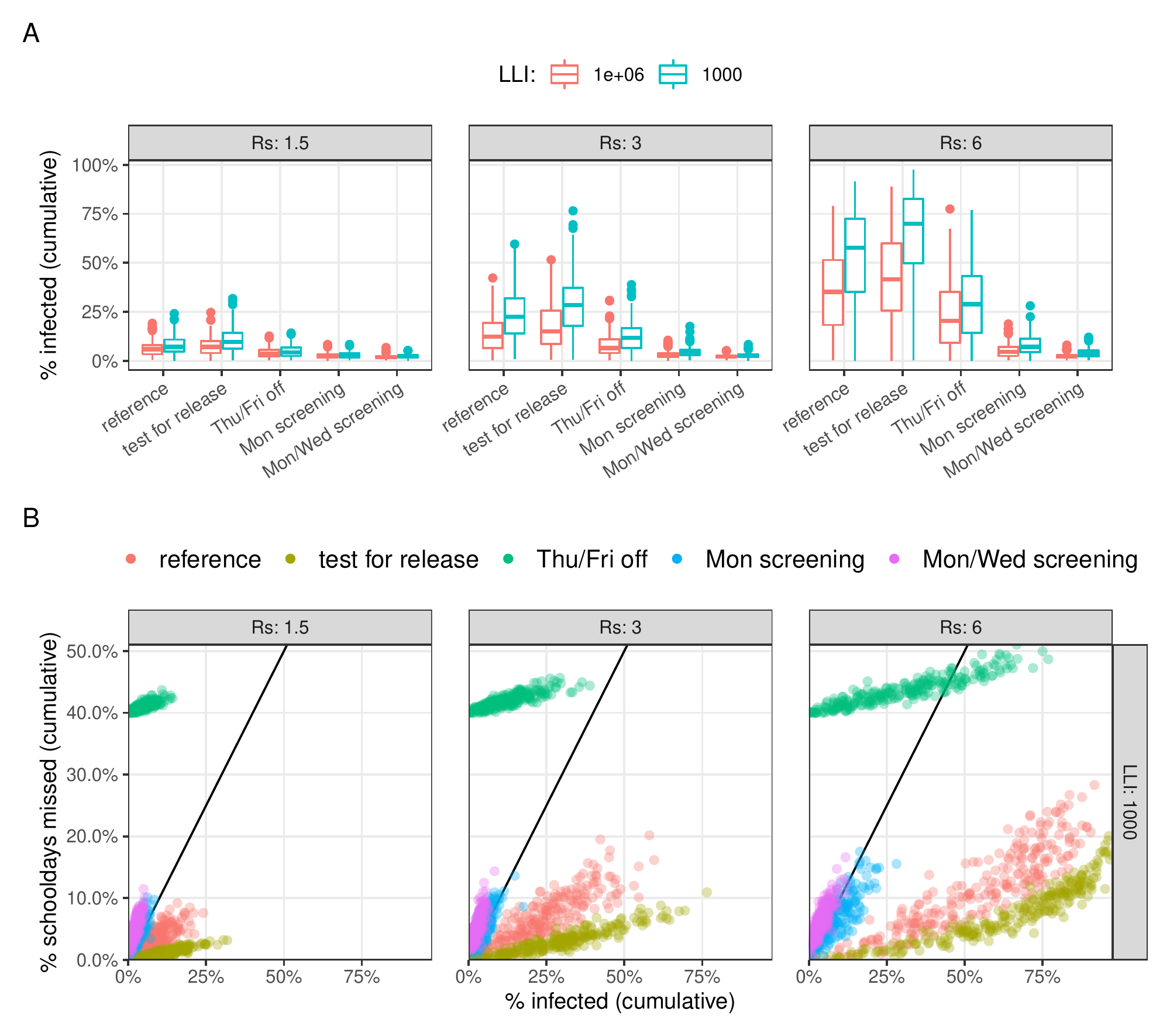}
    \caption{%
        Sensitivity of results with respect to $\lli$;
        $\protect\rzero$ has been re-calibrated for each setting.
    }
    \label{fig:sensitivity-lli}
\end{figure}

A critical factor determining the effectiveness of LFD-test-based policies is
the ratio of test sensitivity relative to the infection probability per risk-contact.
If test-sensitivity is high before individuals show symptoms or have a substantial 
probability of infecting others, 
it is easier to detect asymptomatic cases and contain outbreaks. 
\textit{Vice versa}, a larger limit of infectivity or worse operating characteristics of
an LFD leads to longer time windows of transmitting the virus during the pre- or 
even asymptomatic phase (see Figure~\ref{fig:sensitivity-vs-infectivity}).
We investigate the impact of lowering $\lli$ from $10^6$ (original value proposed in \citep{larremore_test_2021})
to $\lli=10^3$.
To allow for a fair comparison, we re-calibrate $\gamma$ to match the target $\rzero$ values again
(see Figure~\ref{fig:calibration-infectivity}).
This approach allows a more targeted comparison of the relative performance of policies 
with respect to \emph{when} infections occur while keeping the overall level of `infectiousness'
at a comparable level.
Detailed results for this scenario are shown in Figure~\ref{fig:sensitivity-lli}.

The overall structure and relative performance characteristics remain unchanged although
containment of outbreaks is impeded due to the earlier potential to infect others.
However, this affects all policies to some extent, 
irrespective of whether or not they make use of LFD tests.
Twice weekly asymptomatic screening tests in addition to the reference policy of symptomatic bubble
isolation is still able to contain outbreaks fairly effectively.

\subsubsection{Further Scenarios}

We explore two further technical scenarios to assess how the introduction of heterogeneity 
on different levels of the model affects results.
We considered a scenario where additional variation of the VL trajectories was introduced by
adding a temporally correlated Student's t process to the sampled log-10 VL trajectories of 
the Larremore model (see Appendix~\ref{apx:sensitivity-heavy-tails} for details and results).
Alternatively, we explored how between-individual heterogeneity with respect to LFD-test
sensitivity would affect outcome by adding a random effect to equation~\eqref{eq:sensitivity-submodel}
governing the LFD test sensitivity (see Appendix~\ref{apx:sensitivity-random-effect} for details and results).

Under both perturbations the results were remarkably stable and the relative performance 
of the respective policies remained stable.

\section{Discussion} %%%%%%%%%%%%%%%%%%%%%%%%%%%%%%%%%%%%%%%%%%%%%%%%%%%%%%%%%%%%%%%%%

A first and important step to mitigate the impact of schools on the overall infection rate is to control the child-to-child transmission within the school, and this is the question that we addressed in this paper. As there is currently no recommendation of vaccination for children or young persons less than 18 years of age, controlling school outbreaks will remain an important goal in the middle term.

Any model necessarily has to simplify and the choice of modelling tool is dictated by the focus of the analysis at hand.
Other agent-based simulation tools are available and were used to simulate policy impact during Covid-19 outbreaks.
However, these models tend to focus on larger-scale settings \citep{silva2020,li2021} or
local geo-spacial aspects of transmission \citep{vermeulen2020}.
The tool \texttt{openABM}~\citep{hinch2020,openabm_2021} allows the evaluation of very flexible NPIs, 
including delayed reaction to tests and allows agent-based simulations on much larger scale than single schools.
However, for our application, \texttt{openABM} is not tailored to the very fine-grained control required to implement the `test for release' approach and the detailed model for LFD-test sensitivity as function of viral load.
Our agent-based simulation has been set-up carefully to capture important features of the SARS-CoV-2 infection process and how they bear on LFD test results. It has been specifically adapted to the contact structure in schools and has considered a range of policies that have been discussed in the UK or abroad. 
While we have focused our attention on policy implications in schools, 
we stress that our agent-based simulation implementation is flexible and can be adapted 
to the contact structure of other small population environments.
This is relevant to, e.g. workplace environments where the 
contact structure reflects different patterns of workplace interactions such as 
contacts in open plan areas, corridors, or meeting spaces. 
We plan to implement such an extension in future work. 

We stress that we have based our work on the model for viral load presented in \citep{larremore_test_2021}.
This model has been criticised by \citep{deeks_2021} as being unrealistically light-tailed.
We addressed this criticism with extensive sensitivity analyses.
Despite the difficulty of fitting more complex models due to a lack of high-quality data,
our extreme sensitivity scenarios demonstrate that the results obtained are stable across a wide
spectrum of parameter configurations. 
Additionally, these explorations allow deeper insights into the driving factors of successful policies.

Despite a very different approach to modelling the relationship between infectivity and test sensitivity
we reach the same conclusion as \citet{Leng2021} with respect to a dynamic testing regime without 
preemptive isolation of close contacts: testing alone is not sufficient to contain new outbreaks.

The recently released school policy \citep{department_for_education_2021b} recommends repeated testing.
We have taken a simple approach to model compliance, allowing for overdispersion. 
While some data are available, 
compliance patters under repeated testing policies are still largely speculative.
It will thus be important to track and characterise compliance, 
so that in the future realistic modelling of compliance can be calibrated against data.
We do not distinguish between self-testing at home (as currently planned in the UK)
and supervised testing before attending schools.

Further aspects that we did not look into may be of importance when considering the impact of policies in the context of 
school re-openings. 
For instance, the potential effect on within-household transmission from children being at school or adult work-days gained from children being at school. Moreover, we have not considered any potential behavioural impact of a false negative test on the contact pattern of pupils. There has been some discussion of this as a potential issue, but behavioural modelling is beyond the scope of our work. 

Despite the limitations posed by a lack of detailed longitudinal data to fit more complex joint models of 
viral load, infectivity, and test-sensitivity we reach the following conclusions:
\begin{enumerate}
    
    \item Policies cannot be judged on either their ability to contain outbreaks or the amount of face-to-face schooling that they enable alone. Performance can only be judged by considering these quantities jointly and by taking test-burden into account.
    
    \item Depending on the scenario, the distribution of the outcomes of interest may be heavy tailed and simple mean comparison may fail to capture adequately the risks associated with a particular policy.

    \item We found that the relative performance of different policies is qualitatively stable over a wide range of scenarios.
        In particular, additional autocorrelation between repeated testing, lower LFD-test compliance, or a worse $\lli$ profile for infectivity
        all impede outbreak control to some degree but do not change the relative merits and disadvantages of the policies considered.
    
    \item Containment depends on the fraction of asymptomatic cases - it is harder to control outbreaks in scenarios with fewer symptomatic cases. 
        Policies making use of regular asymptomatic screening tests (Mon or Mon/Wed) are generally less affected by this. 
        `Test for release', however, still needs a symptomatic index case to trigger
        dynamic testing within a bubble and thus struggles to contain outbreaks in scenarios 
        with high infectivity and a high fraction of asymptomatic cases. 
        Hence it is a misconception to think that using repeated LFD tests of close contacts as designed in the `test for release' policy is more effective than the reference symptom-based Test~\&~Trace policy when there is a large fraction of asymptomatics. 

    \item Additional autoregression of repeated test results impacts frequent testing performance
        negatively. In particular, the performance of `test for release' in conjunction with 
        low or medium sensitivity screening tests deteriorates. 
        Depending on the time window over which repeated test results are assumed to be 
        correlated, in extreme cases, increased autocorrelation can negate the benefits of testing more than
        once per week. Since no data are available to inform plausible level of additional autocorrelation, our
        results remain simply indicative.
        The additional autocorrelation would however have to be fairly strong to negate the added benefit from a second regular screening day per week.
    
    \item If no effective between-bubble isolation is possible (one bubble per class),
        containment is impeded since the higher number of contacts offsets the wider scope
        of isolation and testing. 
        
    \item The `test for release' policy consistently achieves slightly worse containment than the reference policy at a smaller loss in schooldays.
        Both fare badly in terms of their absolute ability to contain outbreaks however.
        
    \item An extended weekend strategy can only be recommended as a last-resort 
        if no screening tests are available whatsoever, since already a once-weekly 
        regular screening test dominates it clearly. 
    
     \item A once-weekly screening test in addition to symptomatic bubble isolation is already effective.
        A second test per week increases robustness in high-infectivity scenarios. 
        
    \item We conclude that LFD tests are not fit to replace symptomatic isolation of close contacts 
        but that the addition of asymptomatic testing to an existing valid policy shows at least some benefit across all scenarios considered.
        This finding remains valid even if the test sensitivity is fairly low but the degree of additional
        benefit scales with the test quality.
    
\end{enumerate}
We believe that our results have delivered new quantitative understanding of school policy effectiveness for controlling transmission of SARS-CoV-2, and should be used by policy makers to guide the choice of effective policies to be trialled and evaluated, so that schools can stay open for the benefit of our children and their future.

\subsection*{Acknowledgements} %%%%%%%%%%%%%%%%%%%%%%%%%%%%%%%%%%%%%%

We thank Professor Jon Deeks for his helpful comments that lead to
our including the sensitivity analysis with respect to the role of $\lli$.

\subsection*{Funding} %%%%%%%%%%%%%%%%%%%%%%%%%%%%%%%%%%%%%%

Sylvia Richardson's work was funded by the UK Medical Research Council programme\newline\texttt{MRC\_MC\_UU\_00002/10}
and the Alan Turing Institute fellowship \texttt{TU/B/000092}. 

\subsection*{Declaration of Conflicts of Interest} %%%%%%%%%%%%%%%%%%%%%%%%%%%%%%%%%%%%%%

\textit{Sheila Bird:} member of RSS COVID-19 Taskforce; chairs RSS Panel on Test~\&~Trace; member of Testing Initiatives Evaluation Board (since Jan 2021).

\noindent\textit{Sylvia Richardson:} President of the Royal Statistical Society, co-chair of RSS COVID-19 Taskforce; member of the International Best Practice Advisory Group; member of the Joint Biosecurity Centre Data Science Advisory Board.

\FloatBarrier
\Urlmuskip=0mu plus 1mu\relax
\bibliography{references}

\begin{thebibliography}{50}
\providecommand{\natexlab}[1]{#1}
\providecommand{\url}[1]{\texttt{#1}}
\expandafter\ifx\csname urlstyle\endcsname\relax
  \providecommand{\doi}[1]{doi: #1}\else
  \providecommand{\doi}{doi: \begingroup \urlstyle{rm}\Url}\fi

\bibitem[Arons et~al.(2020)Arons, Hatfield, Reddy, Kimball, James, Jacobs,
  Taylor, Spicer, Bardossy, Oakley, et~al.]{arons_2020}
M.~M. Arons, K.~M. Hatfield, S.~C. Reddy, A.~Kimball, A.~James, J.~R. Jacobs,
  J.~Taylor, K.~Spicer, A.~C. Bardossy, L.~P. Oakley, et~al.
\newblock Presymptomatic sars-cov-2 infections and transmission in a skilled
  nursing facility.
\newblock \emph{{N}ew {E}ngland journal of medicine}, 382\penalty0
  (22):\penalty0 2081--2090, 2020.

\bibitem[Baggio et~al.(2020)Baggio, L'Huillier, Yerly, Bellon, Wagner, Rohr,
  Huttner, Blanchard-Rohner, Loevy, Kaiser, Jacquerioz, and
  Eckerle]{baggio_sars-cov-2_2020}
S.~Baggio, A.~G. L'Huillier, S.~Yerly, M.~Bellon, N.~Wagner, M.~Rohr,
  A.~Huttner, G.~Blanchard-Rohner, N.~Loevy, L.~Kaiser, F.~Jacquerioz, and
  I.~Eckerle.
\newblock {SARS}-{CoV}-2 viral load in the upper respiratory tract of children
  and adults with early acute {COVID}-19.
\newblock \emph{Clinical Infectious Diseases: An Official Publication of the
  Infectious Diseases Society of America}, Aug. 2020.
\newblock ISSN 1537-6591.
\newblock \doi{10.1093/cid/ciaa1157}.

\bibitem[Bezanson et~al.(2017)Bezanson, Edelman, Karpinski, and
  Shah]{Julia2017}
J.~Bezanson, A.~Edelman, S.~Karpinski, and V.~B. Shah.
\newblock Julia: A fresh approach to numerical computing.
\newblock \emph{SIAM {R}eview}, 59\penalty0 (1):\penalty0 65--98, 2017.
\newblock \doi{10.1137/141000671}.

\bibitem[Bird et~al.(2005)Bird, Sir~David, Farewell, Harvey, Tim, and
  Peter~C.]{bird_2005}
S.~M. Bird, C.~Sir~David, V.~T. Farewell, G.~Harvey, H.~Tim, and S.~Peter~C.
\newblock Performance indicators: good, bad, and ugly.
\newblock \emph{Journal of the Royal Statistical Society: Series A (Statistics
  in Society)}, 168\penalty0 (1):\penalty0 1--27, Jan. 2005.
\newblock ISSN 0964-1998, 1467-985X.
\newblock \doi{10.1111/j.1467-985X.2004.00333.x}.
\newblock URL \url{http://doi.wiley.com/10.1111/j.1467-985X.2004.00333.x}.

\bibitem[{Children’s Task and Finish
  Group}(2020)]{childrens_task_and_finish_group_childrens_2020}
{Children’s Task and Finish Group}.
\newblock Children’s {Task} and {Finish} {Group}: update to 4th {Nov} 2020
  paper on children, schools and transmission, Dec. 2020.
\newblock URL
  \url{https://www.gov.uk/government/publications/tfc-children-and-transmission-update-paper-17-december-2020}.

\bibitem[Deeks et~al.(2021)Deeks, Gill, Bird, Richardson, and
  Ashby]{deeks_2021}
J.~Deeks, M.~Gill, S.~Bird, S.~Richardson, and D.~Ashby.
\newblock Covid-19 {INNOVA} testing in schools: don’t just test, evaluate,
  Jan. 2021.
\newblock URL \url{http://blogs.bmj.com/bmj/?p=49314}.

\bibitem[{Department for
  Education}(2021{\natexlab{a}})]{department_education_2021}
{Department for Education}.
\newblock Coronavirus ({COVID}-19) asymptomatic testing in schools and
  colleges, 2021{\natexlab{a}}.
\newblock URL
  \url{https://www.gov.uk/government/publications/coronavirus-covid-19-asymptomatic-testing-in-schools-and-colleges/coronavirus-covid-19-asymptomatic-testing-in-schools-and-colleges}.

\bibitem[{Department for
  Education}(2021{\natexlab{b}})]{department_for_education_2021b}
{Department for Education}.
\newblock Schools coronavirus ({COVID-19}) operational guidance,
  2021{\natexlab{b}}.
\newblock URL
  \url{https://assets.publishing.service.gov.uk/government/uploads/system/uploads/attachment_data/file/964351/Schools_coronavirus_operational_guidance.pdf}.

\bibitem[{Department for
  Education}(2021{\natexlab{c}})]{department_for_education_restricting_2021}
{Department for Education}.
\newblock Restricting attendance during the national lockdown: schools,
  2021{\natexlab{c}}.
\newblock URL
  \url{https://assets.publishing.service.gov.uk/government/uploads/system/uploads/attachment_data/file/958906/Restricting_attendance_during_the_national_lockdown_schools_guidance.pdf}.

\bibitem[{Department of Health and Social Care}(2020)]{carehome_LDF_2020}
{Department of Health and Social Care}.
\newblock Evidence summary for lateral flow devices (lfd) in relation to care
  homes, Dec. 2020.
\newblock URL
  \url{https://www.gov.uk/government/publications/evidence-on-the-accuracy-of-lateral-flow-device-testing/evidence-summary-for-lateral-flow-devices-lfd-in-relation-to-care-homes}.

\bibitem[{Department of Health and Social Care}(2021{\natexlab{a}})]{dhsc2021}
{Department of Health and Social Care}.
\newblock Weekly statistics for rapid asymptomatic testing in {England}: 4
  {March} to 10 {March} 2021, 2021{\natexlab{a}}.
\newblock URL
  \url{https://www.gov.uk/government/publications/weekly-statistics-for-nhs-test-and-trace-england-4-march-to-10-march-2021/weekly-statistics-for-rapid-asymptomatic-testing-in-england-4-march-to-10-march-2021}.

\bibitem[{Department of Health and Social
  Care}(2021{\natexlab{b}})]{workplacetesting_2020}
{Department of Health and Social Care}.
\newblock Government boost to rapid workplace testing, Feb. 2021{\natexlab{b}}.
\newblock URL
  \url{https://www.gov.uk/government/news/government-boost-to-rapid-workplace-testing}.

\bibitem[Dinnes et~al.(2020)Dinnes, Deeks, Adriano, Berhane, Davenport,
  Dittrich, Emperador, Takwoingi, Cunningham, Beese, Dretzke, Ruffano, Harris,
  Price, Taylor-Phillips, Hooft, Leeflang, Spijker, Bruel, and
  Group]{dinnes_rapid_2020}
J.~Dinnes, J.~J. Deeks, A.~Adriano, S.~Berhane, C.~Davenport, S.~Dittrich,
  D.~Emperador, Y.~Takwoingi, J.~Cunningham, S.~Beese, J.~Dretzke, L.~F.~d.
  Ruffano, I.~M. Harris, M.~J. Price, S.~Taylor-Phillips, L.~Hooft, M.~M.
  Leeflang, R.~Spijker, A.~V.~d. Bruel, and C.~C.-. D. T.~A. Group.
\newblock Rapid, point‐of‐care antigen and molecular‐based tests for
  diagnosis of {SARS}‐{CoV}‐2 infection.
\newblock \emph{Cochrane Database of Systematic Reviews}, \penalty0 (8), 2020.
\newblock ISSN 1465-1858.
\newblock \doi{10.1002/14651858.CD013705}.
\newblock URL
  \url{https://www.cochranelibrary.com/cdsr/doi/10.1002/14651858.CD013705/full}.
\newblock Publisher: John Wiley \& Sons, Ltd.

\bibitem[FDA(2020)]{fda_2020}
FDA.
\newblock {SARS}-{CoV}-2 {Reference} {Panel} {Comparative} {Data}.
\newblock Dec. 2020.
\newblock URL
  \url{https://www.fda.gov/medical-devices/coronavirus-covid-19-and-medical-devices/sars-cov-2-reference-panel-comparative-data}.
\newblock Publisher: FDA.

\bibitem[Ferretti et~al.(2020)Ferretti, Wymant, Kendall, Zhao, Nurtay,
  Abeler-D{\"o}rner, Parker, Bonsall, and Fraser]{ferretti2020}
L.~Ferretti, C.~Wymant, M.~Kendall, L.~Zhao, A.~Nurtay, L.~Abeler-D{\"o}rner,
  M.~Parker, D.~Bonsall, and C.~Fraser.
\newblock Quantifying {SARS-CoV-2} transmission suggests epidemic control with
  digital contact tracing.
\newblock \emph{Science}, 368\penalty0 (6491), 2020.

\bibitem[Fraser(2021)]{fraser_2021}
C.~Fraser.
\newblock {NHS} {Test and Trace} performance tracker, Feb. 2021.
\newblock URL
  \url{https://www.health.org.uk/news-and-comment/charts-and-infographics/nhs-test-and-trace-performance-tracker}.

\bibitem[{GOV.uk}(2021)]{gov_covid_test_2021}
{GOV.uk}.
\newblock Get a free nhs test to check if you have coronavirus, Feb. 2021.
\newblock URL \url{https://www.gov.uk/get-coronavirus-test}.

\bibitem[GOV.UK(2021)]{govuk_schools_2021}
GOV.UK.
\newblock Schools, pupils and their characteristics, academic year 2019/20,
  2021.
\newblock URL
  \url{https://explore-education-statistics.service.gov.uk/find-statistics/school-pupils-and-their-characteristics}.

\bibitem[Haseltine(2021)]{haseltine_2021}
W.~A. Haseltine.
\newblock Self-testing: A route to school re-opening – the {Austrian}
  example.
\newblock \emph{Forbes}, Feb. 2021.
\newblock URL
  \url{https://www.forbes.com/sites/williamhaseltine/2021/02/26/self-testing-a-route-to-school-re-opening--the-austrian-example/?sh=772f4b0f2d58}.

\bibitem[He et~al.(2021)He, Guo, Mao, and Zhang]{he_2021}
J.~He, Y.~Guo, R.~Mao, and J.~Zhang.
\newblock Proportion of asymptomatic coronavirus disease 2019: A systematic
  review and meta-analysis.
\newblock \emph{Journal of medical virology}, 93\penalty0 (2):\penalty0
  820--830, 2021.

\bibitem[He et~al.(2020)He, Lau, Wu, Deng, Wang, Hao, Lau, Wong, Guan, Tan,
  et~al.]{he_2020}
X.~He, E.~H. Lau, P.~Wu, X.~Deng, J.~Wang, X.~Hao, Y.~C. Lau, J.~Y. Wong,
  Y.~Guan, X.~Tan, et~al.
\newblock Temporal dynamics in viral shedding and transmissibility of covid-19.
\newblock \emph{Nature medicine}, 26\penalty0 (5):\penalty0 672--675, 2020.

\bibitem[Hinch et~al.(2020)Hinch, Probert, Nurtay, Kendall, Wymatt, Hall,
  Lythgoe, Cruz, Zhao, Stewart, et~al.]{hinch2020}
R.~Hinch, W.~J. Probert, A.~Nurtay, M.~Kendall, C.~Wymatt, M.~Hall, K.~Lythgoe,
  A.~B. Cruz, L.~Zhao, A.~Stewart, et~al.
\newblock Openabm-covid19-an agent-based model for non-pharmaceutical
  interventions against covid-19 including contact tracing.
\newblock \emph{medRxiv}, 2020.

\bibitem[Hippich et~al.(2020)Hippich, Holthaus, Assfalg, Zapardiel-Gonzalo,
  Kapfelsperger, Heigermoser, Haupt, Ewald, Welzhofer, Marcus,
  et~al.]{hippich2020}
M.~Hippich, L.~Holthaus, R.~Assfalg, J.~Zapardiel-Gonzalo, H.~Kapfelsperger,
  M.~Heigermoser, F.~Haupt, D.~A. Ewald, T.~C. Welzhofer, B.~A. Marcus, et~al.
\newblock A public health antibody screening indicates a 6-fold higher
  sars-cov-2 exposure rate than reported cases in children.
\newblock \emph{Med}, 2020.

\bibitem[Jones et~al.(2020)Jones, Mühlemann, Veith, Biele, Zuchowski, Hofmann,
  Stein, Edelmann, Corman, and Drosten]{jones_analysis_2020}
T.~C. Jones, B.~Mühlemann, T.~Veith, G.~Biele, M.~Zuchowski, J.~Hofmann,
  A.~Stein, A.~Edelmann, V.~M. Corman, and C.~Drosten.
\newblock An analysis of {SARS}-{CoV}-2 viral load by patient age.
\newblock \emph{medRxiv}, page 2020.06.08.20125484, June 2020.
\newblock \doi{10.1101/2020.06.08.20125484}.
\newblock URL
  \url{https://www.medrxiv.org/content/10.1101/2020.06.08.20125484v1}.
\newblock Publisher: Cold Spring Harbor Laboratory Press.

\bibitem[Kmietowicz(2021)]{kmietowicz2021a}
Z.~Kmietowicz.
\newblock Covid-19: Controversial rapid test policy divides doctors and
  scientists, 2021.

\bibitem[Kunzmann et~al.(2021{\natexlab{a}})Kunzmann, Lingj{\ae}rde, Bird, and
  Richardson]{kunzmann2021a}
K.~Kunzmann, C.~Lingj{\ae}rde, S.~Bird, and S.~Richardson.
\newblock {\texttt{cov19sim}}, 2021{\natexlab{a}}.
\newblock URL \url{https://github.com/kkmann/cov19sim}.
\newblock DOI: 10.5281/zenodo.4623744.

\bibitem[Kunzmann et~al.(2021{\natexlab{b}})Kunzmann, Lingj{\ae}rde, Bird, and
  Richardson]{kunzmann2021b}
K.~Kunzmann, C.~Lingj{\ae}rde, S.~Bird, and S.~Richardson.
\newblock {Supplemental Material: Code for Simulation and Plots},
  2021{\natexlab{b}}.
\newblock URL \url{https://github.com/kkmann/covid-19-screening-policies}.
\newblock DOI: 10.5281/zenodo.4623739.

\bibitem[Larremore et~al.(2021)Larremore, Wilder, Lester, Shehata, Burke, Hay,
  Tambe, Mina, and Parker]{larremore_test_2021}
D.~B. Larremore, B.~Wilder, E.~Lester, S.~Shehata, J.~M. Burke, J.~A. Hay,
  M.~Tambe, M.~J. Mina, and R.~Parker.
\newblock Test sensitivity is secondary to frequency and turnaround time for
  {COVID}-19 screening.
\newblock \emph{Science Advances}, 7\penalty0 (1):\penalty0 eabd5393, Jan.
  2021.
\newblock ISSN 2375-2548.
\newblock \doi{10.1126/sciadv.abd5393}.
\newblock URL \url{https://advances.sciencemag.org/content/7/1/eabd5393}.

\bibitem[Leng et~al.(2021)Leng, Hill, Thompson, Tildesley, Keeling, and
  Dyson]{Leng2021}
T.~Leng, E.~M. Hill, R.~N. Thompson, M.~J. Tildesley, M.~J. Keeling, and
  L.~Dyson.
\newblock Assessing the impact of secondary school reopening strategies on
  within-school covid-19 transmission and absences: a modelling study.
\newblock \emph{medRxiv}, 2021.
\newblock \doi{10.1101/2021.02.11.21251587}.
\newblock URL
  \url{https://www.medrxiv.org/content/early/2021/02/12/2021.02.11.21251587}.

\bibitem[Lennard et~al.(2021)Lennard, Rozmanowski, Pang, Charlett, Anderson,
  Hughes, Barnard, Peto, Vipond, Sienkiewicz, Hopkins, Bell, Crook, Gent,
  Walker, Eyre, and Peto]{peto_2021}
L.~Lennard, S.~Rozmanowski, M.~Pang, A.~Charlett, C.~Anderson, G.~Hughes,
  M.~Barnard, L.~Peto, R.~Vipond, A.~Sienkiewicz, S.~Hopkins, J.~Bell,
  D.~Crook, N.~Gent, S.~Walker, D.~Eyre, and T.~Peto.
\newblock An observational study of {SARS}-{CoV}-2 infectivity by viral load
  and demographic factors and the utility lateral flow devices to prevent
  transmission, 2021.
\newblock URL
  \url{http://modmedmicro.nsms.ox.ac.uk/wp-content/uploads/2021/01/infectivity_manuscript_20210119_merged.pdf}.

\bibitem[Li and Giabbanelli(2021)]{li2021}
J.~Li and P.~J. Giabbanelli.
\newblock Returning to a normal life via covid-19 vaccines in the usa: a
  large-scale agent-based simulation study.
\newblock \emph{medRxiv}, 2021.

\bibitem[Mina et~al.(2021)Mina, Peto, García-Fiñana, Semple, and
  Buchan]{mina_2021}
M.~J. Mina, T.~E. Peto, M.~García-Fiñana, M.~G. Semple, and I.~E. Buchan.
\newblock Clarifying the evidence on {SARS}-{CoV}-2 antigen rapid tests in
  public health responses to {COVID}-19.
\newblock \emph{The Lancet}, 0\penalty0 (0), Feb. 2021.
\newblock ISSN 0140-6736, 1474-547X.
\newblock \doi{10.1016/S0140-6736(21)00425-6}.
\newblock URL
  \url{https://www.thelancet.com/journals/lancet/article/PIIS0140-6736(21)00425-6/abstract}.
\newblock Publisher: Elsevier.

\bibitem[Munday et~al.(2021)Munday, Jarvis, Gimma, Wong, van, and
  Covid]{munday_estimating_2021}
J.~D. Munday, C.~I. Jarvis, A.~Gimma, K.~L. Wong, K.~van, and C.~Covid.
\newblock Estimating the impact of reopening schools on the reproduction number
  2 of {SARS}-{CoV}-2 in {England}, using weekly contact survey data, 2021.
\newblock URL
  \url{https://cmmid.github.io/topics/covid19/reports/comix/schools/School%20Reopening%20-%20PreprintVersion.pdf}.

\bibitem[{NHS Test \& Trace}(2020)]{nhs_2020}
{NHS Test \& Trace}.
\newblock { COVID-19} national testing programme: Schools \& colleges handbook,
  Dec. 2020.
\newblock URL
  \url{http://www.nationalarchives.gov.uk/doc/open-government-licence/version/3/}.

\bibitem[{Office for National Statistics}(2020)]{SIS_1_2020}
{Office for National Statistics}.
\newblock {COVID-19} schools infection survey round 1, england, Dec. 2020.
\newblock URL
  \url{https://www.ons.gov.uk/peoplepopulationandcommunity/healthandsocialcare/conditionsanddiseases/bulletins/covid19schoolsinfectionsurveyround1england/november2020}.

\bibitem[Oran and Topol(2020)]{oran_2020}
D.~P. Oran and E.~J. Topol.
\newblock Prevalence of asymptomatic sars-cov-2 infection: a narrative review.
\newblock \emph{Annals of internal medicine}, 173\penalty0 (5):\penalty0
  362--367, 2020.

\bibitem[{Oxford Big Data Institute: Pathogen Dynamics
  Group}(2021)]{openabm_2021}
{Oxford Big Data Institute: Pathogen Dynamics Group}.
\newblock {BDI}-pathogens/{OpenABM}-{Covid19}, Feb. 2021.
\newblock URL \url{https://github.com/BDI-pathogens/OpenABM-Covid19}.

\bibitem[{R Core Team}(2020)]{r2020}
{R Core Team}.
\newblock \emph{R: A Language and Environment for Statistical Computing}.
\newblock R Foundation for Statistical Computing, Vienna, Austria, 2020.
\newblock URL \url{https://www.R-project.org/}.

\bibitem[SAGE(2020)]{sage-12-2020}
SAGE.
\newblock Seventy-fourth {SAGE} meeting on {COVID-19}, Dec. 2020.
\newblock URL
  \url{https://assets.publishing.service.gov.uk/government/uploads/system/uploads/attachment_data/file/948606/s0991-sage-meeting-74-covid-19.pdf}.

\bibitem[Shah et~al.(2014)Shah, Wilson, and Ghahramani]{shah2014studentt}
A.~Shah, A.~G. Wilson, and Z.~Ghahramani.
\newblock Student-t processes as alternatives to gaussian processes, 2014.

\bibitem[Silva et~al.(2020)Silva, Batista, Lima, Alves, Guimar{\~a}es, and
  Silva]{silva2020}
P.~C. Silva, P.~V. Batista, H.~S. Lima, M.~A. Alves, F.~G. Guimar{\~a}es, and
  R.~C. Silva.
\newblock Covid-abs: An agent-based model of covid-19 epidemic to simulate
  health and economic effects of social distancing interventions.
\newblock \emph{Chaos, Solitons \& Fractals}, 139:\penalty0 110088, 2020.

\bibitem[Sutton et~al.(2020)Sutton, Fuchs, D’alton, and Goffman]{sutton_2020}
D.~Sutton, K.~Fuchs, M.~D’alton, and D.~Goffman.
\newblock Universal screening for sars-cov-2 in women admitted for delivery.
\newblock \emph{New {E}ngland {J}ournal of {M}edicine}, 382\penalty0
  (22):\penalty0 2163--2164, 2020.

\bibitem[{The DELVE Initiative}(2020)]{delve-4}
{The DELVE Initiative}.
\newblock Balancing the risks of pupils returning to schools, 2020.
\newblock URL
  \url{https://rs-delve.github.io/reports/2020/07/24/balancing-the-risk-of-pupils-returning-to-schools.html}.

\bibitem[Thomas et~al.(2021)Thomas, Angrist, Cameron-Blake, Hallas, Kira,
  Majumdar, Petherick, Phillips, Tatlow, and Webster]{oxford-npi}
H.~Thomas, N.~Angrist, E.~Cameron-Blake, L.~Hallas, B.~Kira, S.~Majumdar,
  A.~Petherick, T.~Phillips, H.~Tatlow, and S.~Webster.
\newblock Oxford covid-19 government response tracker, 2021.

\bibitem[{University of Liverpool}(2020)]{liverpool_covid-19_2020}
{University of Liverpool}.
\newblock Liverpool community testing pilot, interim evaluation, 2020.
\newblock URL
  \url{https://www.liverpool.ac.uk/media/livacuk/coronavirus/Liverpool,Community,Testing,Pilot,Interim,Evaluation.pdf}.

\bibitem[Vermeulen et~al.(2020)Vermeulen, Pyka, and M{\"u}ller]{vermeulen2020}
B.~Vermeulen, A.~Pyka, and M.~M{\"u}ller.
\newblock An agent-based policy laboratory for covid-19 containment strategies,
  2020.

\bibitem[Wald et~al.(2020)Wald, Schmit, and Gusland]{wald_2020}
E.~R. Wald, K.~M. Schmit, and D.~Y. Gusland.
\newblock A pediatric infectious disease perspective on covid-19.
\newblock \emph{{C}linical {I}nfectious {D}iseases}, 2020.

\bibitem[Wheale and Adams(2021)]{wheale_2021}
S.~Wheale and R.~Adams.
\newblock English school leaders despair over new rules on {Covid} tests and
  masks.
\newblock \emph{The Guardian}, Feb. 2021.
\newblock URL
  \url{http://www.theguardian.com/education/2021/feb/25/english-school-leaders-despair-at-soft-line-on-covid-tests-and-masks}.

\bibitem[Whittaker(2021)]{whittaker_2021}
F.~Whittaker.
\newblock Secondary schools can start testing pupils on-site before march 8,
  {DfE} confirms.
\newblock \emph{Schools Week}, Feb. 2021.
\newblock URL
  \url{https://schoolsweek.co.uk/secondary-schools-can-start-testing-pupils-on-site-before-march-8-dfe-confirms/}.

\bibitem[Wise(2020)]{wise_covid-19_2020}
J.~Wise.
\newblock Covid-19: {Lateral} flow tests miss over half of cases, {Liverpool}
  pilot data show.
\newblock \emph{BMJ}, 371:\penalty0 m4848, Dec. 2020.
\newblock \doi{10.1136/bmj.m4848}.
\newblock URL \url{http://www.bmj.com/content/371/bmj.m4848}.
\newblock Publisher: British Medical Journal Publishing Group Section: News.

\end{thebibliography}

\clearpage
\appendix
\section{Appendix}
\counterwithin{figure}{section}

\subsection{Contact Matrices}

\begin{figure}[H]
    \centering
    \includegraphics[width=\textwidth]{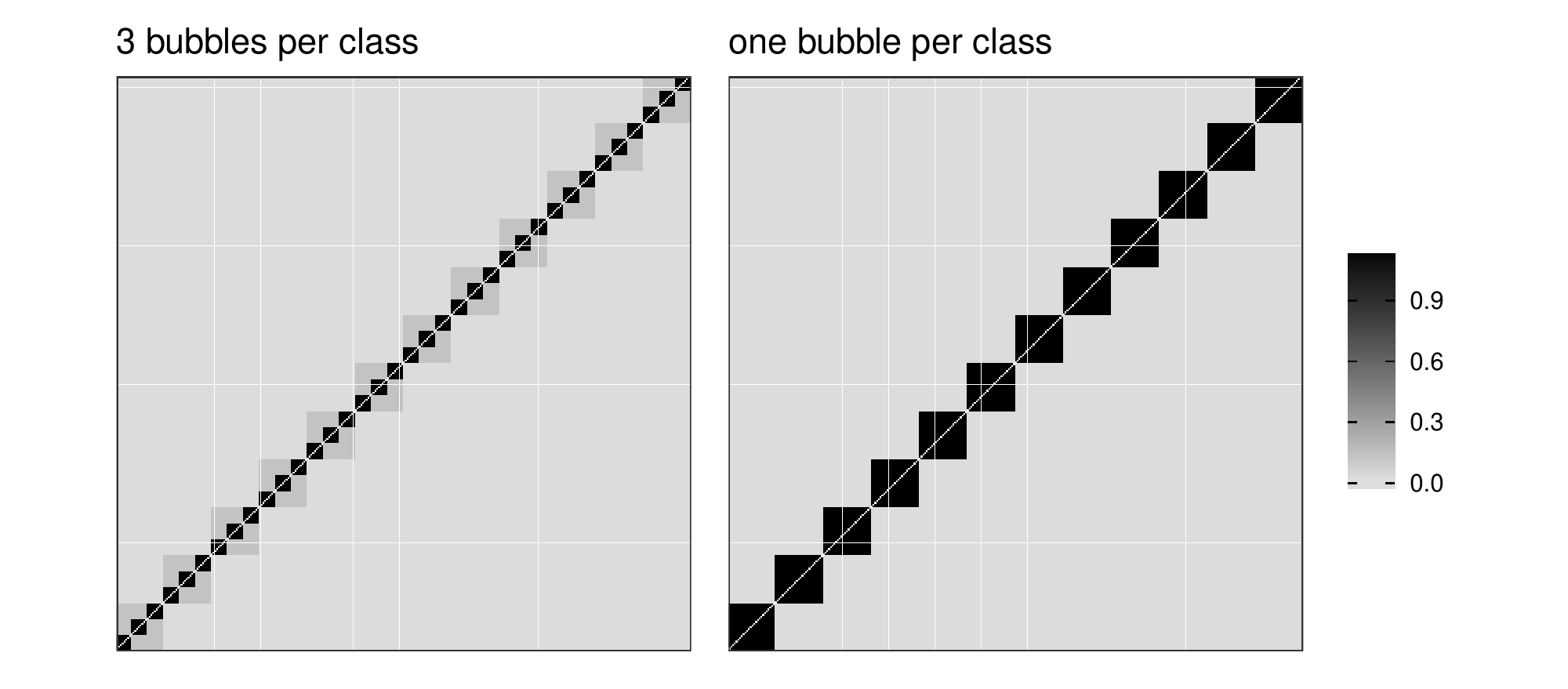}
    \caption{%
        Adjacency matrix of a typical school with either 12 classes of 3 bubbles á 9 pupils each or only one bubble per class respectively.
        Connectivity strength is given in terms of expected number of daily \emph{pair-wise} risk-contacts assuming that
        there is no within-bubble isolation ($p_{\,\operatorname{bubble}}=1$),
        limited between-bubble isolation ($p_{\operatorname{class}} = 3/(|\operatorname{class}| - 1)$), and each pupil has an expected number of
        school-wide contacts of $1$ ($p_{\operatorname{school}} = 1/(|\operatorname{school}| - 1)$).
    }
    \label{fig:adjacency-matrix}
\end{figure}

\subsection{Calibration} %%%%%%%%%%%%%%%%%%%%%%%%%%%%%%%%%%%%%%%%%%%%%%%%%%%%%%%%%%%%%%%%

The proposed overall model requires calibration with respect to to crucial parameters:
We follow \citet{larremore_test_2021} in matching the inactivity constant $\gamma$ 
to the replication number $\rzero$.
The operating characteristics of the screening test are matched to data presented in 
\citet{liverpool_covid-19_2020} and \citet{peto_2021}.

\subsubsection{Infectivity}
\label{sec:calibration-infectivity}

\begin{figure}[H]
    \centering
    \includegraphics[width=\textwidth]{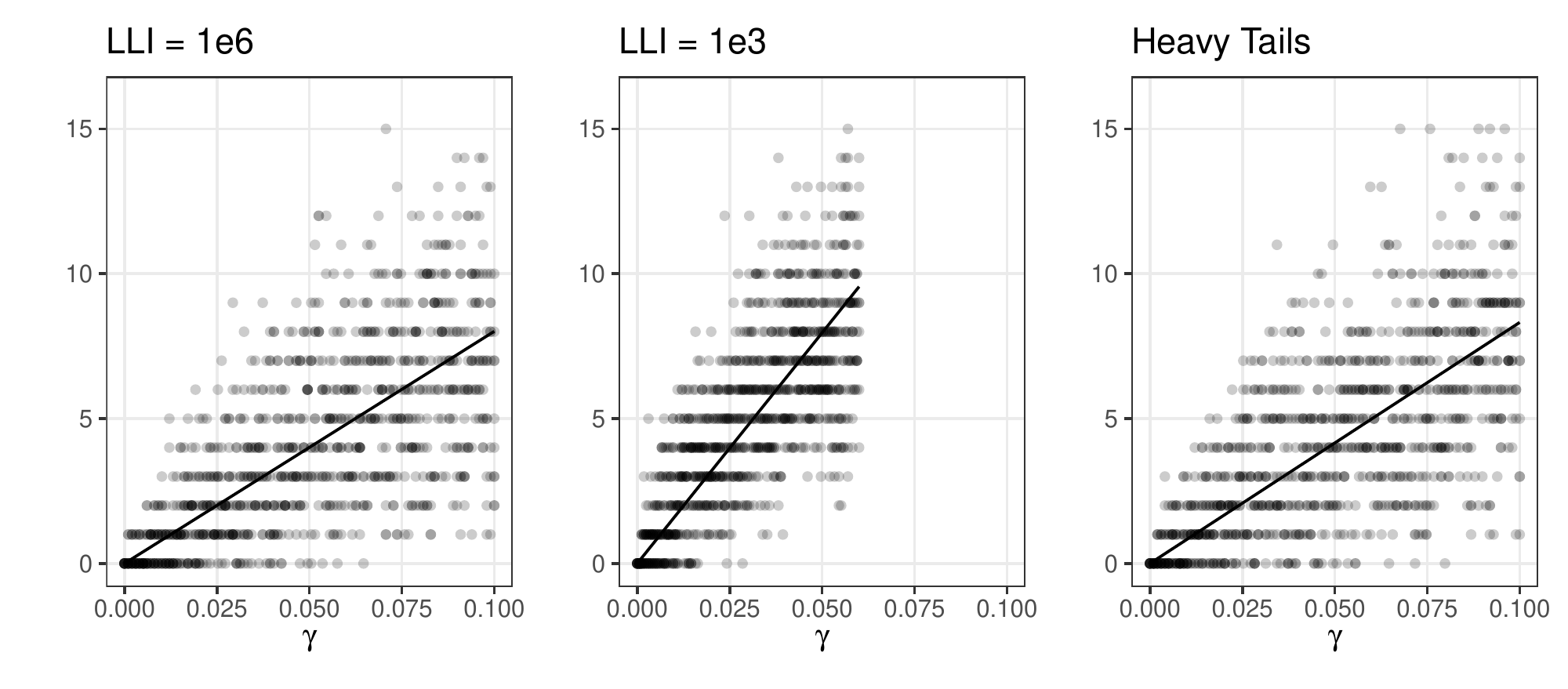}
    \caption{%
        Fitted calibration lines for 1000 simulated populations with 10 resamples 
        for 100 equidistant values of $\gamma$ increasing from $0$ to $0.1$ ($\lli=10^6$, and `Heavy Tails') 
        and $0$ to $0.06$ ($\lli=10^3$).
    }
    \label{fig:calibration-infectivity}
\end{figure}

We simulate forward  for a given model and a given value of $\rzero$ under no policy intervention with a single
index infection at day $0$ and a follow-up of $21$ days.
For each simulation run, the actual reproduction number is determined as the number of 
individuals infected by the index case via exact contact tracing.
To derive the infectivity constant $\gamma$ as a function of the target population $\rzero$, 
we fit a linear regression.
We then use numerical root finding to invert the fitted conditional mean and identify
the $\gamma$ giving rise to a particular $\rzero$.
The calibration does depend on the fraction of asymptomatic cases since their viral
load trajectories are different under the Larremore-model. 
We use a medium value of $50\%$ asymptomatic cases to derive the calibration curves shown in
Figure~\ref{fig:calibration-infectivity}.
This then allows us to derive $\gamma(\rzero)$ for the sensitivity parameter $\rzero$.
Figure~\ref{fig:calibration-infectivity} shows the calibration curves for two different
lower limits of infectivity $\lli=10^6$ (original Larremore~\textit{et~al.}~model) and
$\lli=10^3$ as well as the `Heavy Tails' scenario discussed in Section~\ref{apx:sensitivity-heavy-tails}.

\subsubsection{LFD Test Sensitivity}
\label{sec:calibration-sensitivity}

\begin{figure}[H]
    \centering
    \includegraphics[width=\textwidth]{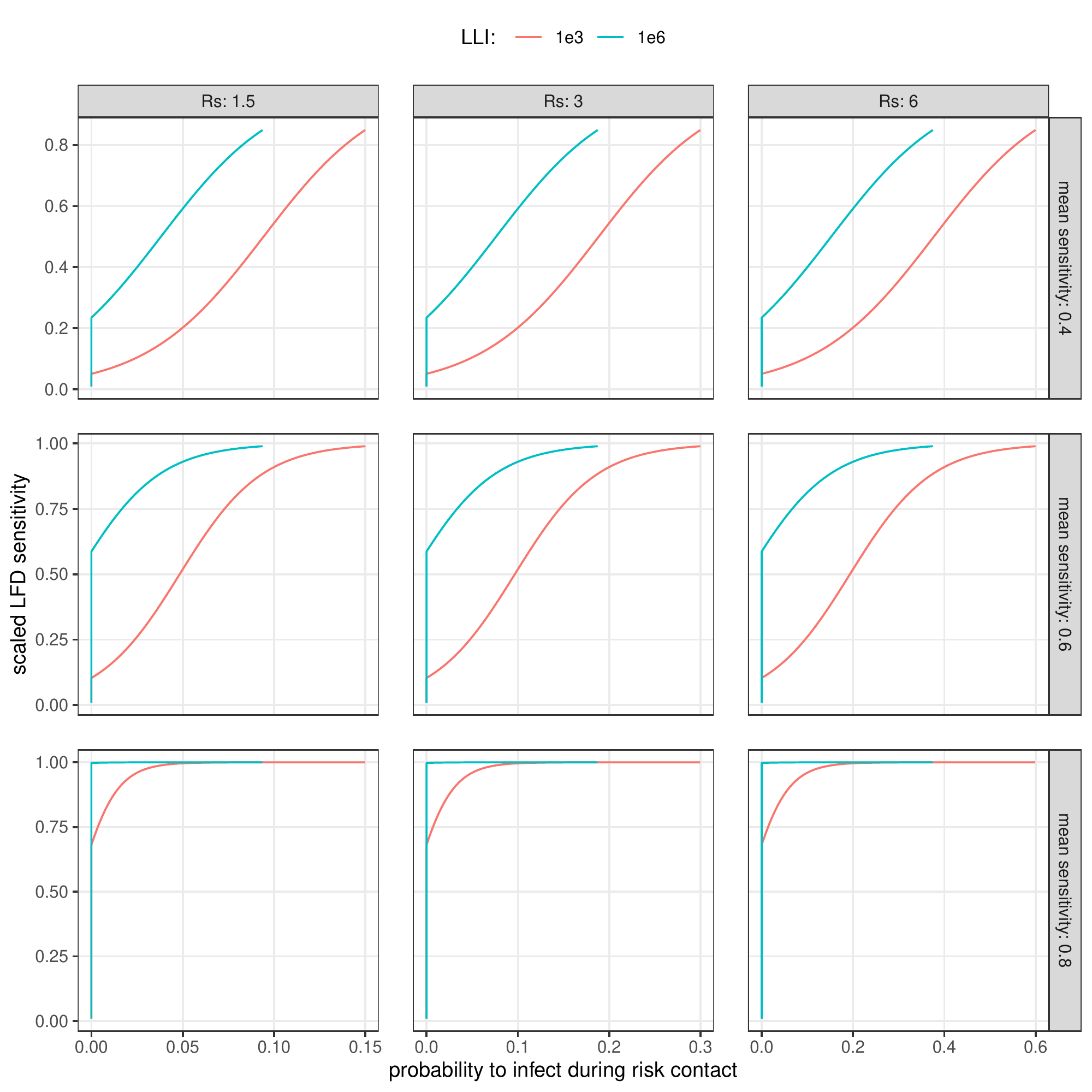}
    \caption{%
        Implied association between infection probability and \emph{scaled} LFD-test sensitivity (equation~\eqref{eq:scaled-sensitivity}) for the~9 scenarios defined in terms of infectivity and \emph{mean} LFD-test sensitivity (`$x$' in equation~\eqref{eq:mean-sensitivity}).
    }.
    \label{fig:sensitivity-vs-infectivity}
\end{figure}

We begin by fitting the logistic regression model~\eqref{eq:sensitivity-submodel} 
to data presented in~\citet{peto_2021} 
to obtain the shape of the relationship between $\vl$ and sensitivity. 
Since we were unable to obtain the raw data, we fit a logistic curve to a set of control
points directly read off the Innova curve in Figure~S1~\citep{peto_2021}.
The fitted model can then be related to data presented by the\citet{liverpool_covid-19_2020}.
The Liverpool pilot found that the test sensitivity of the Innova test in a
practical setting for \emph{pre-symptomatic} individuals was $40\%$ ($95\%$ confidence interval: $28.5\%$ to $52.4\%$)
which is in line with findings in \citet{dinnes_rapid_2020} for other rapid antigen tests.
This information can be used to scale the fitted logistic regression model such that
the mean sensitivity corresponds to the findings of the Liverpool study.
To this end we introduce a scaling factor $\eta$ to reconcile the
shape of the sensitivity curve found in the Oxford data with the mean sensitivity
of the real-world experiment from Liverpool by considering \emph{scaled sensitivity}
\begin{align}
\operatorname{sensitivity}_\eta(\operatorname{VL}) :&= \operatorname{logit}^{-1}\big( 
	\beta_{\operatorname{VL}} \cdot \log_{10}\big( \operatorname{VL}^\eta \big)
	+ c_{\operatorname{test}} \big) \ . \label{eq:scaled-sensitivity}
\end{align}
We simulate $10^5$ viral load trajectories 
(assuming a moderate rate of $50\%$ asymptomatic cases) 
and randomly select one pre-symptomatic viral load value per trajectory
resulting in a cross-sectional sample $\vl_i, i = 1\ldots l, l\leq 10^5$ of viral
load values mimicking the structure of the Liverpool data set.
For any given target mean sensitivity $x$, the final value of $\eta$ is then identified by solving
\begin{align}
    \frac{1}{l}\,\sum_{i=1}^m\operatorname{sensitivity}_\eta(\operatorname{VL_i}) = x \label{eq:mean-sensitivity}
\end{align}
for $\eta$.
We explore three sensitivity scenarios ($x = 0.4, x = 0.6$, and $x = 0.8$) in the main simulation 
study.
\begin{figure}[H]
    \centering
    \includegraphics[width=\textwidth]{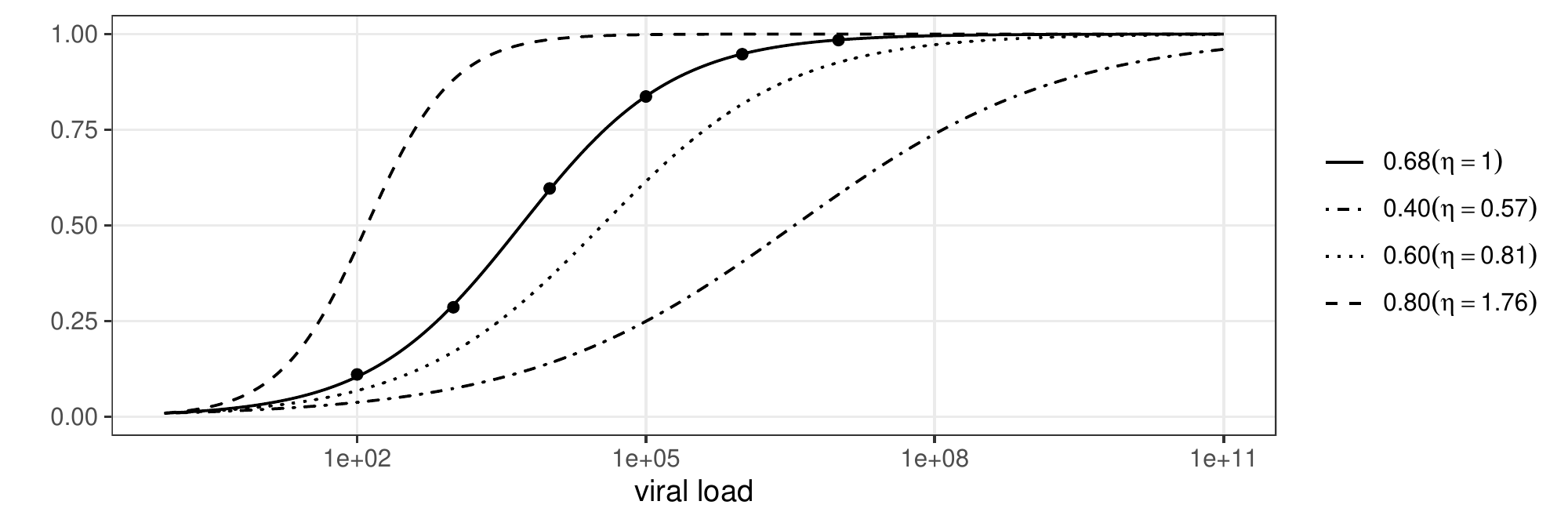}
    \caption{%
        Fitted sensitivity model with control points for the Innova LFD test 
        ($\eta=1$, implied mean pre-symptomatic sensitivity of $0.82$) 
        and the scaled models corresponding to implied mean pre-symptomatic sensitivities of 
        $0.4, 0.6$, and $0.8$.
    }.
    \label{fig:calibration-test-scaled-sensitivity}
\end{figure}
A crucial property of the overall model is the implied relationship between 
the infection probability and LFD test sensitivity.
This is induced by their joint dependency on the latent viral load trajectories.
Since we consider three scenarios for infectivity ($\rzero=1.5, 3, 6$) and test sensitivity
(sensitivity of $0.4, 0.6, 0.8$) each, 
this implies 9 scenarios of the dependency between infection probability and test sensitivity.
Additionally, we consider a scenario where $\lli=1000$ instead of $\lli=10^6$ as in \citep{larremore_test_2021}
(see Figure~\ref{fig:sensitivity-vs-infectivity} and Section~\ref{sec:sensitivity-lli} for results).

\begin{figure}[H]
    \centering
    \includegraphics[width=\textwidth]{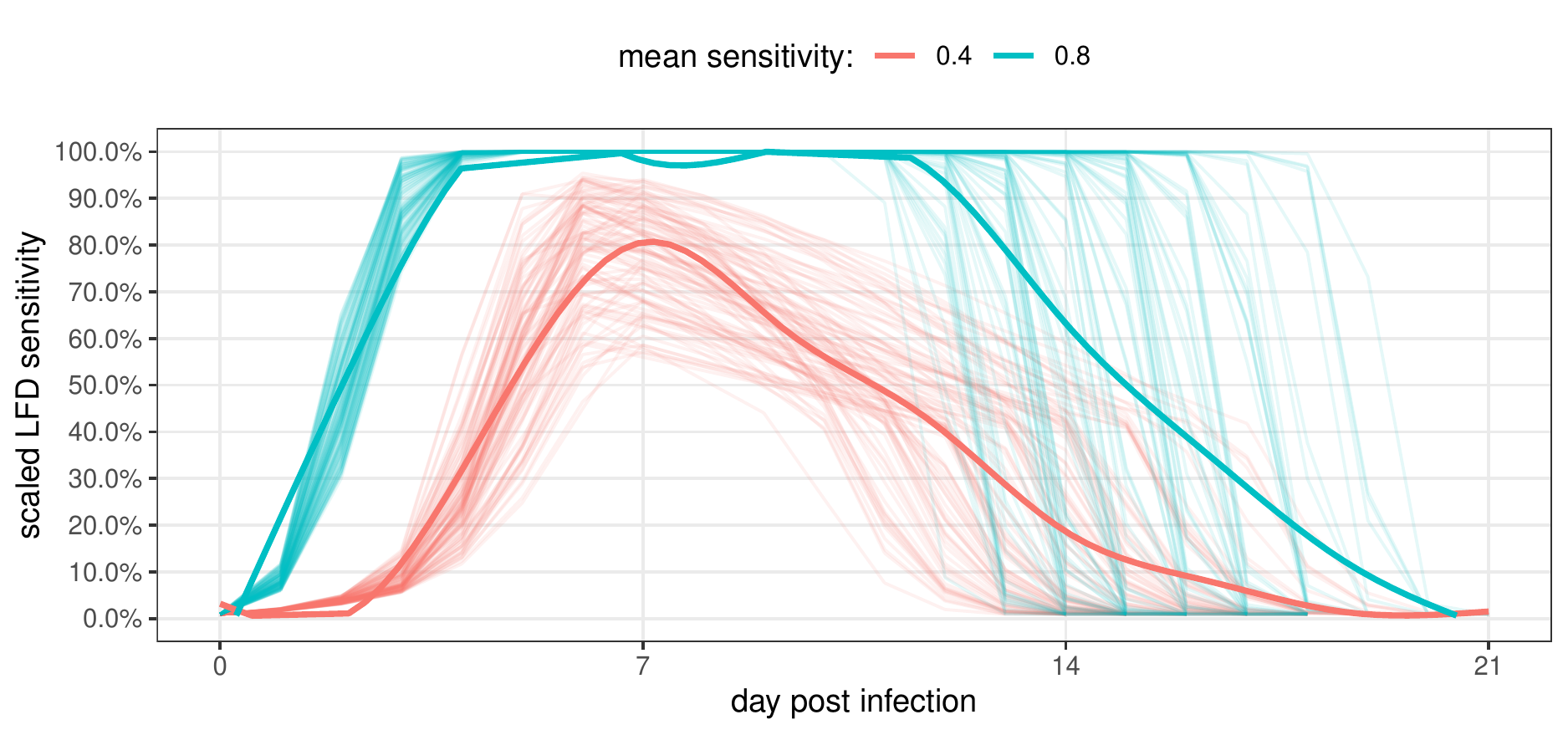}
    \caption{%
        Sample of 100 trajectories of test sensitivity over time for two most extreme mean sensitivity scenarios; all other parameters correspond to the baseline scenario.
    }.
    \label{fig:temporal-shift-test-sensitivity}
\end{figure}

The temporal shift of varying LFD sensitivity is visualised in 
Figure~\ref{fig:temporal-shift-test-sensitivity}.
The main impact of higher test sensitivity is that infected individuals can be identified earlier in time.

\clearpage
\subsection{LFD Test Compliance}
\label{apx:compliance}

\begin{figure}[H]
    \centering
     \includegraphics[width=\textwidth]{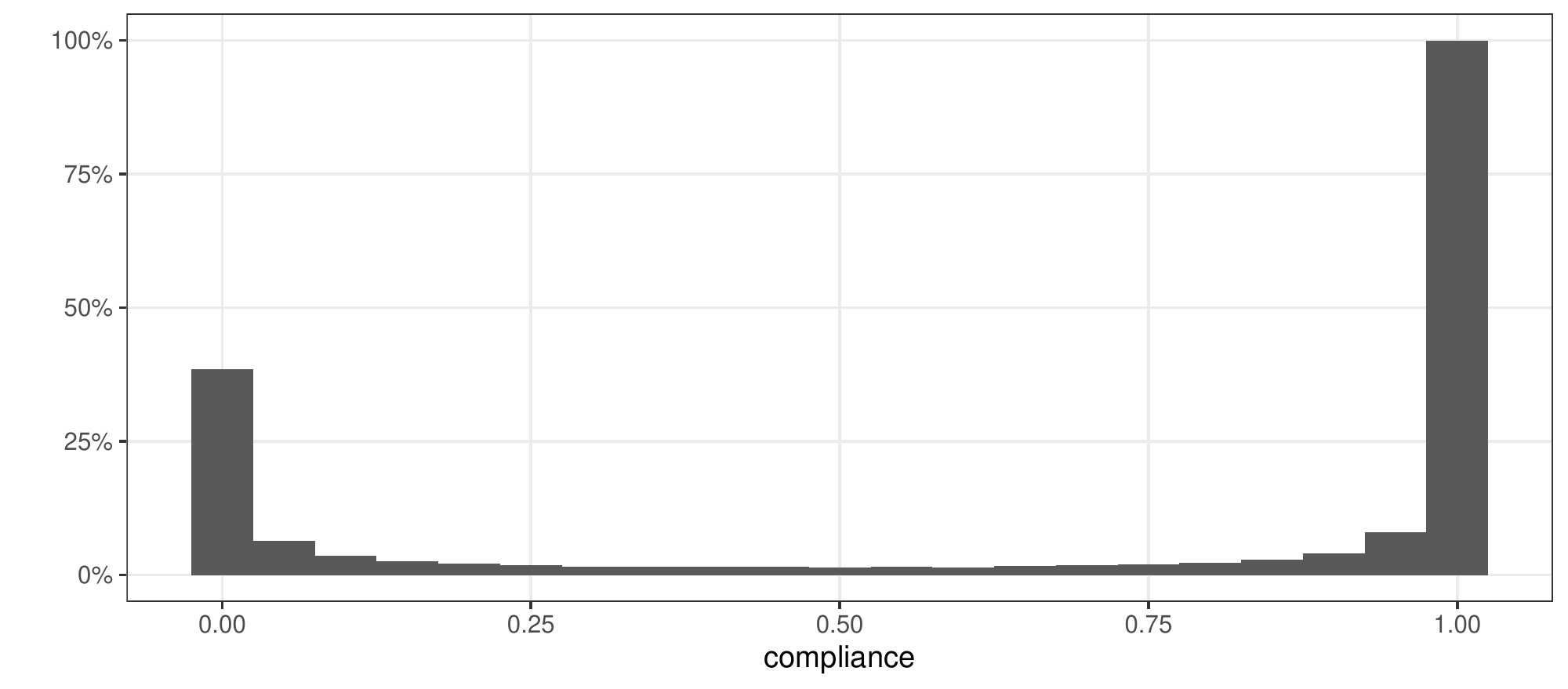}
     \caption{%
        Histogram of the Beta distribution used to sample the individual LFD-test compliance probabilities.
    }
    \label{fig:sensitivity-compliance-beta}
\end{figure}

We model individual compliance with LFD testing by drawing a random effect per-pupil from
a $\operatorname{Beta}(2/15,1/15)$ distribution (see Figure~\ref{fig:sensitivity-compliance-beta}).
This implies a population mean compliance of $66.7\%$.
Whenever an LFD test is required by a policy, an independent biased coin toss is sampled using the 
pupil's compliance probability to determine whether the LFD test is actually conducted or not.
The U-shape was chosen to reflect the assumptions that an individuals choice to comply with
LFD testing will correlate over time.
Compliance with PCR testing is always $100\%$.

\clearpage
\subsection{More Variation in the VL Trajectories} %===================
\label{apx:sensitivity-heavy-tails}

\begin{figure}[H]
    \centering
    \includegraphics[width=\textwidth]{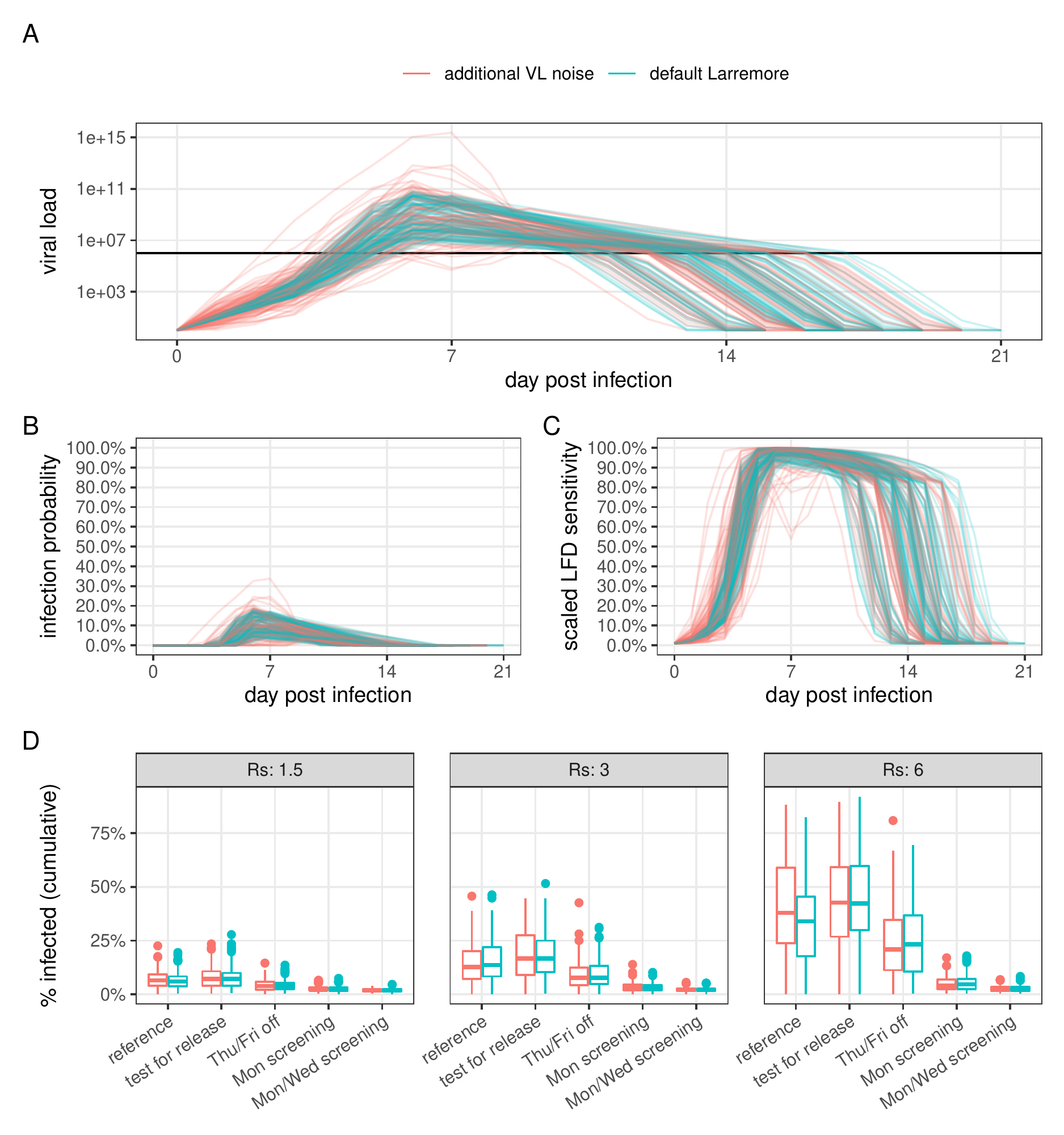}
    \caption{%
        Sensitivity of results with respect to additional heavy-tailed noise on the VL trajectories for the 50\% asymptomatic cases and 60\% mean pre-symptomatic LFD test sensitivity (baseline scenario).
    }
    \label{fig:sensitivity-heavy-tails}
\end{figure}

The model used by \cite{larremore_test_2021} to sample VL trajectories leads to 
smooth trajectories of each individual and the distribution of peak-VL values is
drawn from an uniform distribution on the log-10 scale (see Section~\ref{sec:vl-model}). 
To explore how additional variation in VL, and in particular heavier tails of the
distribution of VL, would affect the results, 
we added correlated, heavy tailed noise to the the trajectories obtained under
the Larremore model.
Specifically, we used a Student's t process \citep{shah2014studentt} with 3 degrees of freedom, 
a squared exponential covariance function with length scale 5, and a scaling
factor of of the noise of $3/\sqrt(3)$ which results in a marginal standard deviation of 3.
We restricted the additional noise to the first 10 days after onset and conditioned the samples
to zero noise at days $0$ and $10$.
The restriction the the first 10 days ensures that the clearance phase is smooth and
that individuals do not switch between being infectious and not being infectious from day 10 onward.
We additionally conditioned the samples on positive trajectories and trajectories with a maximal VL of $10^15$.
Figure~\ref{fig:sensitivity-heavy-tails} shows 100 sample paths drawn under both the default Larremore model
and the model with additional correlated VL noise (panel A),
the corresponding infection probability (panel B), the LFD test sensitivity (panel C),
and the resulting proportion of infected pupils (panel D). 
The infectivity constant $\gamma$ has been calibrated to the respective setting to ensure that the
resulting $\rzero$ values match.

Differences between the two VL models in the policies' ability to contain outbreaks are minimal.
This is due to the fact that the additional variation of VL affects both infectivity and sensitivity -
individuals with high viral load are thus more likely to be LFD positive as well.

\begin{figure}[H]
    \centering
    \includegraphics[width=\textwidth]{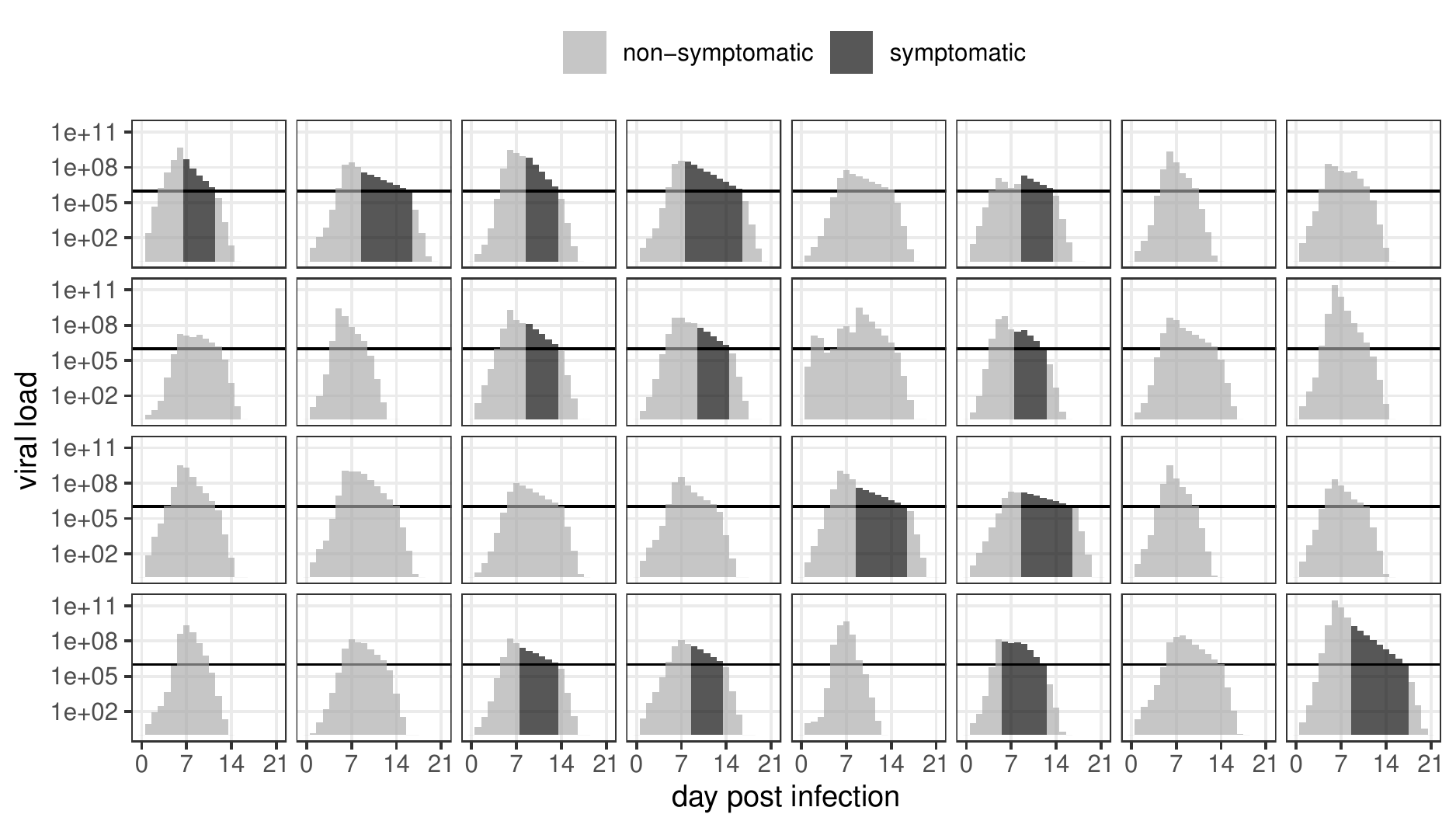}
    \caption{%
        VL trajectories under the Larremore model with added noise.
    }
    \label{fig:vl-trajectories-heavy-tails}
\end{figure}

\subsection{Between-Subject Variation of LFD Sensitivity} %===================
\label{apx:sensitivity-random-effect}

So far, it was assumed that both LFD test sensitivity and infection probability were deterministic
functions of VL.
By varying the sensitivity curves for fixed $\gamma$ the population-mean 
temporal lag between becoming infectious 
and being detectable by LFD can be varied (see Figure~\ref{fig:sensitivity-vs-infectivity}).
However, there might also be heterogeneity between individuals with respect to LFD-sensitivity, 
for example linked to the way they perform the nasal swabs.
In contrast to the scenario investigated in Appendix~\ref{apx:sensitivity-heavy-tails},
this heterogeneity only would affects LFD-test sensitivity and thus decouples the deterministic relationship
between test sensitivity and infection probability.

We model this by adding a normally distributed random effect to equation~\eqref{eq:sensitivity-submodel}
for each individual $i$ 
\begin{align}
    g_i(\vl_t) 
        := \operatorname{logit}^{-1}\big( 
	        \beta_{\operatorname{test}} \, \logten(\vl_t) + \beta_u \,u_i +  c_{\operatorname{test}}
        \big)
\end{align}
where $U_i\sim\mathcal{N}(0,1), iid.$~and the coefficient $\beta_u$ scales the population heterogeneity. 
Note that a non-zero random effect has consequences for the mean pre-symptomatic sensitivity 
due to the non-linearity  of Equation~\eqref{eq:mean-sensitivity}. 
This means that the scaling factor $\eta$ has to be adjusted to match the target mean sensitivity
for each scenario separately. 

\begin{figure}[H]
    \centering
    \includegraphics[width=\textwidth]{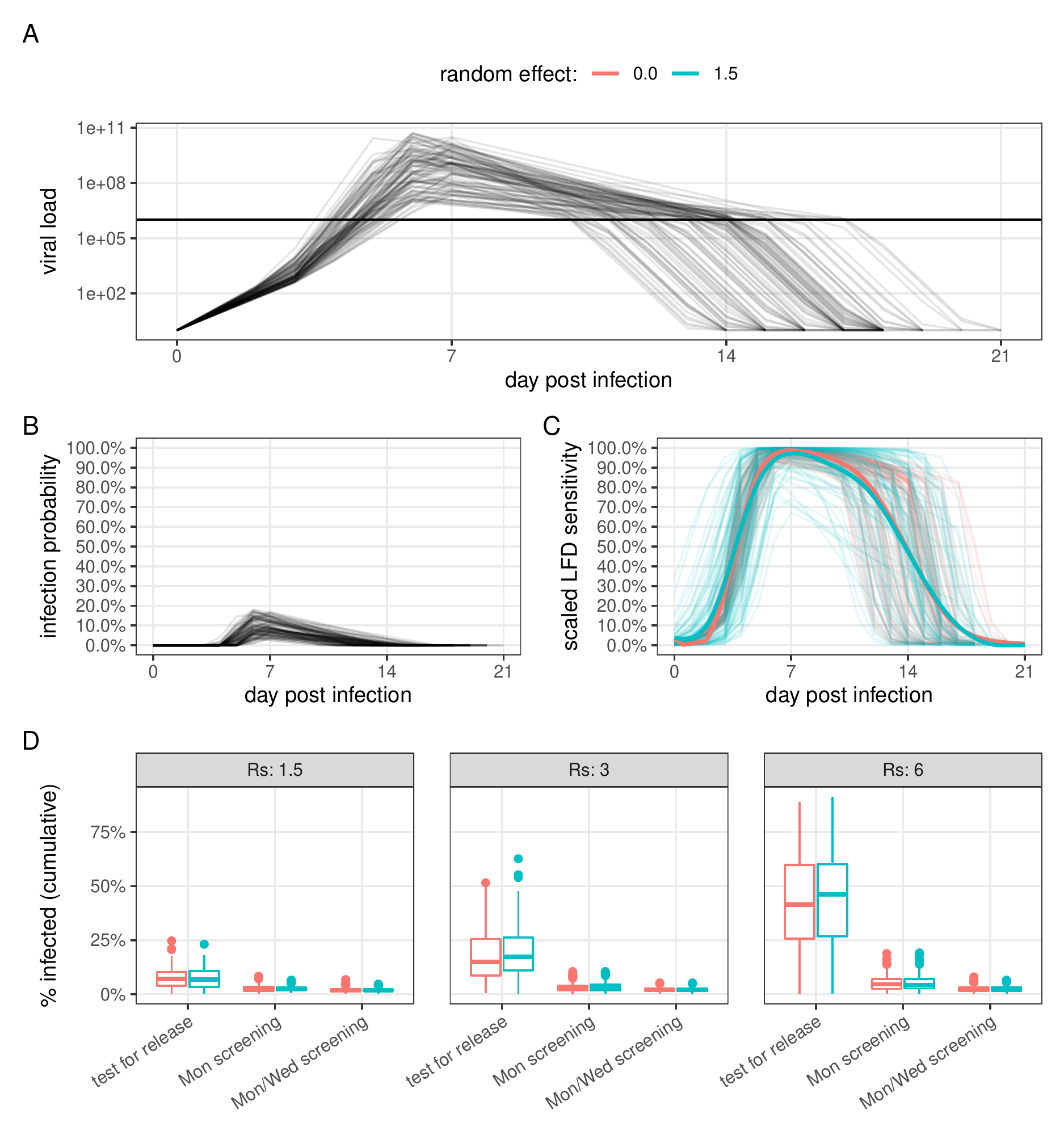}
    \caption{%
        Sensitivity of results with respect between-subject variability of test sensitivity for the 50\% asymptomatic cases and 60\% mean pre-symptomatic LFD test sensitivity (baseline scenario).
    }
    \label{fig:sensitivity-random-effects}
\end{figure}

We see no notable effect of adding moderate heterogeneity on the performance of the policies using LFD tests (see Figure~\ref{fig:sensitivity-random-effects}).

\end{document}